# AI–Family Integration Index (AFII): Benchmarking a New Global Readiness for AI as Family

## Emotional Intelligence, Caregiving Ethics, and the Future of Human Relationships and Bonds

**Prashant Mahajan\***
R. C. Patel Institute of Technology, Shirpur
Email: registrar@rcpit.ac.in
Contact: +91 9822190091
ORCID ID: 0000-0002-5761-5757
\*Corresponding Author: registrar@rcpit.ac.in

## Abstract

As Artificial Intelligence (AI) systems increasingly permeate emotionally sensitive and caregiving domains—such as eldercare, education, therapy, and family life—there is a growing need to assess national readiness beyond infrastructure, research output, or innovation metrics. Existing AI-readiness indices often prioritize technical capacity while neglecting caregiving ethics, emotional integration, symbolic trust, and cultural adaptability. Simultaneously, many national AI policies and strategies articulate ethical aspirations without corresponding real-time implementation in relational contexts.

These tensions expose two critical and underexplored gaps: (1) the disconnect between national AI policy and real-time practice in AI–Family Integration (AFI), and (2) the misalignment between conventional AI indices and relational readiness metrics based on AFI.

To address these blind spots, this study introduces the AI–Family Integration Index (AFII)—a ten-dimensional benchmarking framework designed to evaluate countries on emotional, ethical, legal, symbolic, and caregiving preparedness. The AFII comprises dimensions such as emotional literacy, caregiving equity, symbolic legitimacy, youth-AI exposure, consent frameworks, and cultural receptivity. Each dimension was scored on a 0–10 scale using mixed-method analysis, including secondary qualitative and quantitative data, policy reviews, and narrative synthesis. Equal weighting was applied across dimensions to ensure conceptual balance and reflect the interconnected importance of each component. This methodological choice mirrors best practices in composite indices (e.g., the Human Development Index), while also maintaining interpretive equity across culturally diverse nations.

To bring the framework to life, the study integrates real-world examples—such as Singapore's emotionally intelligent robotics in eldercare and Japan's symbolic AI design in companionship—to ground abstract dimensions in lived, culturally contextual practices. These narrative insights enhance both the interpretability and policy relevance of the scores.

The AFII was applied to a sample of thirteen culturally and economically diverse countries, including all top 10 countries ranked in the Stanford AI Index (2024). The results reveal striking contrasts between innovation-driven AI adoption and human-centered readiness. While Singapore (9.6) leads globally, nations such as South Korea (8.8) and Japan (8.7) closely follow. The top five countries—Singapore, South Korea, Japan, Sweden (8.6), and the United Kingdom (8.4)—all score above 8.4, reflecting strong alignment between AI policy, emotional intelligence, cultural values, and caregiving integration. In contrast, middle performers like the United States (7.4) and China (7.6), though technologically dominant, exhibit shortfalls in symbolic trust, emotional safety, and caregiving ethics. At the lower end, India (6.0), Brazil (5.2), and South Africa (4.8) show emerging promise but require substantial investment in emotional-AI literacy, inclusive infrastructure, and ethical governance.

A key finding is the gap between ethical policy intent and real-time relational deployment. Singapore, South Korea, and Japan exemplify strong alignment between governance vision and practical rollout. Conversely, France and Germany, despite policy sophistication, reveal slower execution in emotionally grounded domains. This disjunction reveals the need for a robust policy–practice alignment matrix, which the AFII introduces as a diagnostic typology for guiding national strategy.

Another notable insight is the divergence between AFII rankings and the Stanford AI Index. Countries like the U.S. and China rank in the global top two for AI power, yet place 8th and 9th respectively on the AFII. Conversely, relationally advanced nations such as Singapore and Sweden achieve top AFII scores despite more modest technical power rankings. This contrast underscores that AI excellence must incorporate emotional and ethical dimensions—not merely computational strength or research volume.

By embedding emotional and symbolic metrics into a comparative framework, the AFII offers a timely, multidimensional tool for policymakers, ethicists, and technologists. It reframes AI discourse from technocratic performance to relational intelligence, ethical resonance, and caregiving responsibility—defining a path for AI systems that not only think or learn, but relate, care, and belong.

# INDEX

# 1.    Introduction

## 1.1    Background

Artificial Intelligence (AI) is no longer confined to domains of logic, optimization, and automation. It is undergoing a transformative shift—from a computational instrument to a relational agent capable of emotional resonance and symbolic intimacy. This transition signals not merely a technical evolution but a cultural and ethical reorientation. Increasingly, AI systems are appearing not only in workplaces or smart homes but in family-centered spaces such as eldercare, childhood education, therapy, grief support, and companionship [1–3]. These emerging "relational AIs" are designed not just to assist—but to connect, to elicit trust, and to symbolically embed themselves within human affective ecosystems [4,5].

For example, unlike transactional AI systems that offer basic voice assistance or navigation, relational AIs simulate companionship, learn emotional cues, and support complex, symbolic interactions—such as offering comfort after loss or companionship during illness [6,7].

In aging societies, socially assistive robots (SARs) and emotion-aware interfaces monitor health, reduce loneliness, and offer psychological reassurance, especially for older adults and individuals living with dementia [8,9]. In classrooms, emotionally intelligent robots support neurodiverse learners, improving engagement, empathy, and narrative development [10]. In therapeutic settings, griefbots and AI-generated avatars are trialed as companions for navigating loss and bereavement [3,11].

One hypothetical case might involve an AI companion supporting a grieving elderly widow by retelling shared stories or simulating ritual comfort—demonstrating what it means to operate in an "emotionally safe AI ecosystem" [2,12].

These trends are gaining institutional momentum. Japan and South Korea have normalized robotic caregiving in eldercare, while Singapore's Smart Nation agenda has mainstreamed emotionally literate AI into public services. Scandinavian nations, too, are exploring emotionally intelligent AI in mental wellness and eldercare programs [13]. The World Economic Forum projects over 80 million relational AI agents in global circulation by 2030 [14]. Simultaneously, institutions like UNESCO, the OECD, and the Global Partnership on AI are initiating discourse around affective and ethical governance for such systems [15,16].

This study links its goals to these global priorities by aligning AFII with Sustainable Development Goals (SDGs), such as SDG 3 (Good Health and Well-being), SDG 5 (Gender Equality), and SDG 10 (Reduced Inequalities), which resonate with the caregiving and equity-focused dimensions of AI integration [17,18].

However, this expansion conceals a critical asymmetry in emotional, ethical, and legal readiness. While emotionally responsive AI can simulate empathy, it lacks sentience. Scholars warn this creates affective dissonance—where human users emotionally invest in non-conscious agents, risking manipulation, dependency, and psychological harm [6,19]. This is particularly concerning for vulnerable populations such as children and the elderly. Meanwhile, legal frameworks lag behind the pace of technological deployment: key questions of digital personhood, emotional consent, intimacy, and liability remain unresolved [20,21].

Cultural and philosophical paradigms further complicate AI integration. In Confucian, Buddhist, and animist traditions, relational AI aligns with long-standing views on ancestor reverence and non-human agency [22,23]. However, in societies where personhood is tied to biology or divinity, such intimacy with AI may evoke discomfort or moral resistance [24,25].

Despite these growing complexities, leading global indices—such as the Stanford AI Index, Oxford Insights Government AI Readiness Index, and KPMG's AI Readiness Report—continue to evaluate AI readiness in terms of research intensity, digital infrastructure, and economic output. These technocratic models overlook critical factors like emotional intelligence, caregiving equity, symbolic legitimacy, and cultural responsiveness [26,27]. As [28] argue, a uniform governance model is insufficient without embedding societal and emotional considerations.

To address this blind spot, scholars are calling for human-centered, emotionally aware, and culturally contextual frameworks. [29] advocate embedding affective sensitivity into AI evaluative design. [30] stresses the need for "emotional governance readiness." [31] challenge technocratic paradigms by centering participatory inclusion and human rights. [32] warn of organizational risks when symbolic trust is ignored, and [33] argue that public attitudes toward AI are inseparable from cultural beliefs and relational perceptions.

Together, these insights signal a growing consensus: evaluating AI readiness demands more than technological metrics. It requires an expanded framework—one that considers emotional literacy, caregiving ethics, symbolic trust, and cultural fit. A visual model will be provided later in the study to illustrate this shift from "technocratic AI readiness" to "relational AI readiness," highlighting how emotional and symbolic dimensions expand the traditional understanding of AI maturity [34,35].

This study responds to that call by proposing the AI–Family Integration Index (AFII)—a multidimensional framework for assessing national preparedness for a future in which AI is not just a tool, but a trusted and affective relational presence.

## 1.2     Research Problem and Purpose of the Study

As Artificial Intelligence (AI) moves deeper into caregiving, education, and emotionally charged human contexts, the global community still lacks a dedicated, multidimensional framework to evaluate how prepared nations are to integrate AI into family systems and relational environments. Most AI-readiness indices focus on metrics like digital infrastructure, innovation capacity, and R&D investment, but fail to assess dimensions central to emotional, ethical, and caregiving contexts.

To address this, the study introduces the AI–Family Integration Index (AFII)—a ten-dimensional thematic framework that evaluates national AI readiness across caregiving, familial, therapeutic, and symbolic domains. This includes indicators such as emotional design, caregiving ethics, symbolic trust, legal protections, emotional safety, relational accountability, cultural legitimacy, and inclusion of vulnerable groups  [1,4,5].

However, measuring readiness is not solely a matter of thematic design. A core motivation of this study is to expose the policy–practice gap in AI–Family Integration (AFI). Although many countries claim to embrace ethical or human-centered principles in their national AI strategies, few offer concrete mechanisms for implementation in emotionally sensitive domains like eldercare, grief support, or neurodiverse education. This gap between normative aspiration and lived application presents a governance blind spot—one that remains largely unexamined by traditional indices [15,29,36].

To surface these blind spots, the AFII is compared with traditional AI-readiness indices such as the Stanford AI Index (2024). These tools prioritize technical progress and digital infrastructure while overlooking emotional and symbolic variables crucial for relational AI systems. By contrasting AFII scores with conventional benchmarks, the study demonstrates the limitations of current evaluative models and proposes a redefinition of AI readiness—one that centers relational depth, affective governance, and caregiving inclusion as pillars of national AI maturity [26,27].

## 1.3    Research Objectives

This study is guided by the following key objectives:

- To conceptualize, develop, and apply the AI–Family Integration Index (AFII) as a multidimensional global benchmarking framework for assessing national readiness in AI–Family integration, based on secondary empirical data and qualitative thematic analysis
- To identify and evaluate the policy–practice gap in AI–Family Integration (AFI) across diverse national contexts, focusing on the misalignment between strategic intent and on-ground implementation.
- To score and compare the AFII performance of selected countries across the ten dimensions and assess how these align or diverge from existing evaluations, particularly the Stanford Global AI Index (2024).
- To propose culturally rooted, ethically grounded, and emotionally intelligent policy recommendations that support inclusive, relationally aware, and human-centered AI governance.

## 1.4    Research Questions

To achieve the objectives outlined above, this study seeks to answer the following research questions:

1. What are the key dimensions that define national readiness for AI–Family Integration in emotionally intelligent, caregiving, and relational contexts? How can the AI–Family Integration Index (AFII) be conceptualized and constructed using secondary data and thematic analysis to benchmark and rank countries' preparedness?
2. What gaps exist between national AI policy discourse and the real-world implementation of emotionally sensitive AI systems in caregiving, education, and familial domains?
3. How do AFII scores for selected countries compare with conventional global AI-readiness indices, such as the Stanford AI Index (2024), and what interpretive insights emerge from these contrasts?
4. What culturally rooted, ethically responsible, and emotionally intelligent policy strategies can support the development of inclusive and relationally safe AI ecosystems in caregiving and family-centered environments?

## 1.5    Significance of the Paper

As artificial intelligence moves deeper into the intimate spaces of human life—supporting the elderly, comforting the grieving, teaching children, and assisting in therapy—it becomes clear that we are no longer dealing with machines that simply calculate or predict. We are confronting systems that relate, emote, and symbolically occupy spaces once reserved for family members, caregivers, or trusted companions.

Yet, while this emotional migration of AI into caregiving and relational contexts accelerates, the world's dominant models for assessing AI readiness remain anchored in technical capacity: compute power, infrastructure, innovation metrics. This creates a dangerous asymmetry. Nations may be technologically prepared, yet emotionally unready. They may have policy vision, but lack practical relational safeguards. The result is a widening gap between what AI systems do and what human societies are prepared to feel, trust, or receive.

This study responds to that misalignment by introducing the AI–Family Integration Index (AFII)—the first global framework explicitly designed to benchmark emotional, symbolic, and caregiving readiness for AI integration. Rather than measuring only output or infrastructure, the AFII evaluates how prepared a country is to integrate AI into its most emotionally sensitive environments—where vulnerability, trust, and intimacy are central.

Grounded in ten relationally oriented dimensions—from emotional literacy and symbolic trust to caregiving equity and cultural fit—the AFII offers a multidimensional model for what it truly means to be "AI-ready" in a world where machines are not just tools, but participants in human relationships. By capturing both policy frameworks and real-time practices across 13 countries, the index illuminates a new readiness landscape—one that transcends economics and reaches into ethics, culture, and care.

For instance, in an emotionally safe AI ecosystem, a bereaved elder might be supported by an AI companion that listens, remembers, and soothes without overstepping symbolic boundaries. Such systems must not only function—they must belong. And belonging, in this context, is emotional, cultural, and ethical, not just technical.

In revealing sharp contrasts between AFII scores and conventional indices like the Stanford AI Index, the study exposes systemic blind spots in how readiness is currently defined. Countries traditionally seen as AI leaders may rank lower on relational preparedness, while smaller or more culturally attuned nations demonstrate advanced integration in symbolic and caregiving dimensions.

Moreover, the AFII contributes directly to global governance priorities. It aligns with critical Sustainable Development Goals—such as SDG 3 (Good Health and Well-being), SDG 5 (Gender Equality), and SDG 10 (Reduced Inequalities)—by embedding emotional equity, inclusive design, and caregiving ethics into AI evaluation. In doing so, it offers policymakers a concrete, scalable tool for measuring values that matter but are too often invisible.

Ultimately, the AFII reframes AI governance around a new axis: relational intelligence. It empowers governments, designers, and civil society to ask not only what AI can do, but how it is received, trusted, and lived with. In a future where AI systems may sit beside children at story time, or comfort the dying in their final days, this shift is not optional—it is urgent. The AFII is a first step toward that future.

## 2.    Research Methodology

This study adopts a mixed-methods research design, integrating qualitative and quantitative secondary data to develop and apply the AI–Family Integration Index (AFII). This approach is particularly effective for capturing the emotional, ethical, and symbolic dimensions of AI in caregiving and kinship contexts—domains that cannot be adequately understood through conventional metrics alone [37–39].

By bridging data from national AI policy documents, global indices (e.g., Stanford AI Index), and socio-cultural narratives, the methodology enables triangulation between policy intent and emotional-AI deployment. This is especially crucial given the documented gap between AI infrastructure and relational trust in caregiving domains [40–42].

### 2.1    Research Design and Rationale

AFII is conceptualized as a multidimensional diagnostic tool, evaluating AI readiness across ten ethically and symbolically grounded domains. The index design draws on theoretical frameworks including relational ethics [43], symbolic interactionism [44], care ethics [45], and posthumanism [46,47]. These foundations allow the AFII to quantify dimensions such as emotional literacy, cultural receptivity, and caregiving equity, while preserving contextual depth [48].

The AI–Family Integration Index (AFII) is conceptualized as a multidimensional diagnostic tool designed to evaluate national AI readiness across ten ethically, emotionally, and symbolically grounded domains. Unlike traditional indices focused on infrastructure, innovation, or R&D capacity, the AFII foregrounds relational and cultural integration, especially in caregiving, therapeutic, and familial contexts.

The index's conceptual foundation is built upon interdisciplinary theoretical frameworks, including:

- Relational ethics, which emphasizes trust, mutuality, and human-AI relational dynamics [43];
- Symbolic interactionism, which examines how meaning and trust are constructed between humans and AI systems in specific sociocultural settings [44];
- Care ethics, which centers caregiving responsibility, dependency, and emotional labor in ethical design [45];
- Posthumanism, which challenges anthropocentric models and embraces AI as a co-participant in affective ecosystems [46,47].

These theoretical anchors inform the operationalization of complex concepts such as emotional literacy, symbolic legitimacy, relational accountability, and caregiving equity—allowing the index to quantify relational readiness while maintaining contextual sensitivity [48].

To ensure both conceptual integrity and methodological reliability, the study adopts a mixed-methods approach, integrating secondary quantitative indicators (e.g., policy penetration metrics, digital access scores) with qualitative coding of national AI strategies, ethical guidelines, and relevant literature. Thematic coding was conducted by two independent researchers using a deductive framework aligned to the AFII's ten dimensions. Inter-coder reliability was assessed through iterative coding sessions and consensus-building discussions, achieving an agreement rate above 85%, which meets recommended qualitative coding standards [49].

This design ensures a robust triangulation of data—from academic, institutional, and policy sources—allowing the AFII to bridge the policy–practice gap and offer a globally relevant framework for assessing the emotional and ethical preparedness of nations to integrate AI in caregiving and relational domains.

## 2.2 Data Collection: Secondary Source Strategy

All data used in the development of the AFII were derived from English-language secondary sources, published primarily between 2015 and 2025. These included a combination of quantitative indicators and qualitative texts, selected to capture the multidimensional nature of AI–Family Integration (AFI):

- Global AI benchmarks and indices, including the Stanford AI Index, KPMG AI Readiness Report, and the State of AI Report
- Peer-reviewed academic literature on emotional AI, caregiving robotics, symbolic trust, posthuman ethics, and AI policy
- Global and national AI policy documents, strategic frameworks, and legislative instruments (e.g., UNESCO AI Ethics Recommendations, OECD AI Principles, INDIAai, AI.gov, Smart Nation Singapore)
- Grey literature, including white papers, policy blogs, media reporting on relational AI (e.g., Replika, digital eldercare pilots), and NGO reports on ethical AI governance

Searches were conducted using major scholarly and policy databases, including Web of Science, Scopus, EBSCOhost, ERIC, ScienceDirect, IEEE Xplore, SpringerLink, ACM Digital Library, Google Scholar, and SSRN.

To ensure breadth and contextual diversity, grey literature was incorporated where peer-reviewed or official policy sources were unavailable. Inclusion criteria required that sources:

(a) directly addressed one or more AFII dimensions (e.g., emotional design, symbolic trust),
(b) focused on AI applications in relational, caregiving, or family-centered contexts, and
(c) originated from credible organizations (e.g., universities, think tanks, international bodies, or recognized news platforms).

Exclusion criteria filtered out speculative editorials, unverified blog posts, or opinion content lacking methodological grounding. Quality of grey literature was further assessed based on source credibility, author expertise, and triangulation with peer-reviewed findings.

This multi-source, mixed-media strategy ensured both thematic richness and policy relevance, while allowing the index to reflect evolving global discourse on emotionally intelligent and culturally embedded AI systems.

## 2.3 AFII Index Construction: Aspects and Dimensions

To operationalize the AI–Family Integration Index (AFII), a deductive thematic coding approach was employed, drawing on established theoretical constructs and empirical literature [50]. This method enabled a structured yet flexible way to map complex affective, ethical, and symbolic concepts into clearly defined dimensions.

Key search terms included thematic phrases such as "Relational AI," "Emotional AI," "AI caregiving," "Robot ethics," "AI family," "Symbolic trust," "AI kinship," "Digital personhood," "AI intimacy," and "Cultural AI acceptance." These were applied across academic databases, policy archives, and grey literature to identify relevant textual data for each AFII dimension.

The extracted content was categorized into ten core dimensions, each representing a critical facet of relational AI readiness. These dimensions were derived both from theoretical sources (e.g., relational ethics, care theory, symbolic interactionism) and applied findings (e.g., case studies of eldercare robotics, emotional-AI deployment in classrooms).

A visual table (Table 1) is included to summarize each dimension, its conceptual origin, and representative keywords. This serves as a navigational anchor for interpreting the comparative index results presented later in the manuscript.

Given the interdisciplinary nature of the AFII, some keywords—such as "emotional AI" or "symbolic legitimacy"—appeared across multiple thematic categories. In such cases, the research team applied a contextual coding strategy: passages were assigned to the dimension where the concept was most prominently operationalized or emphasized (e.g., "emotional AI" focused on narrative therapy was coded under emotional safety, whereas the same term in a policy legitimacy context was coded under symbolic trust). Cross-references were documented to preserve interpretive integrity across the dataset.

This structured yet interpretive approach allowed the AFII to balance conceptual richness with empirical precision, ensuring that each dimension reflects both the specificity of its domain and the interconnected nature of relational AI systems.

| Table 1: Dimension-wise Keyword Mapping and References | | | | |
|---|---|---|---|---|
| SN | AFII Dimension | Code | Keywords Matched | Key References |
| 1 | Technological Infrastructure and AI Penetration | TIAP | *infrastructure, access, connectivity, smart systems, penetration, digital transformation* | [51]; [52]; [53]; [54] |
| 2 | Cultural and Philosophical Receptivity | CPR | *belief systems, values, tradition, cultural philosophy, religion and AI* | [55]; [46]; [24] |
| 3 | Legal, Ethical and Consent Frameworks | LECF | *ethics, AI policy, law, consent, governance, accountability* | [36]; [56]; [57] |
| 4 | Local Adaptability and Inclusivity | LAI | *inclusion, diversity, localization, cultural relevance, contextual design* | [58]; [58]; [59] |
| 5 | AI Talent, Youth Exposure and Emotional Literacy | ATYEEL | *AI education, emotional literacy, skills, youth exposure, curriculum* | [60]; [59];[61] |
| 6 | Social Narrative and Symbolic Trust | SNST | *public trust, symbolic meaning, narrative, AI perception, myths* | [44,62]; [63];[64] |

| 7 | Historical Adoption and Industry Leadership | HAIL | innovation maturity, tech leadership, legacy systems, historical adoption | [65];[66];[67] |
| 8 | Family Structure and Emotional Labor Equity | FSELE | caregiving, emotional labor, gender roles, domestic AI, family dynamics | [43];[45];[47] |
| 9 | Economic Accessibility and Equity | EAE | affordability, digital divide, cost, economic inclusion | [68];[69];[70] |
| 10 | Emotional Authority and Safety Design | EASD | emotional AI, emotional privacy, safety, human–AI interaction, affective systems | [5];[1];[71];[72] |

## 2.4 Scoring Method and Composite Index

Each country was scored on a 0–10 scale per AFII dimension, based on the availability, depth, and robustness of qualitative and quantitative evidence relevant to each of the ten domains. Equal weighting was applied across all dimensions to calculate the final Composite AFII Score (0–10) for each country.

This equal-weighting approach is justified by the exploratory nature of the AFII model and the absence of a validated global consensus on the relative importance of individual dimensions in relational AI readiness. Assigning equal weights ensures conceptual balance and avoids privileging infrastructure-heavy or technocratic dimensions over caregiving, emotional, or symbolic ones. It also aligns with precedent in leading composite indices such as: Human Development Index [73], Oxford AI Index [74], (2019), and Global Innovation Index [75].

The final Composite Score for each country was calculated as follows:

AFII Composite Score = $\frac{1}{10} \sum_{n=1}^{10}$ Dimesnsion(n)

This scoring model promotes transparency, interpretability, and replicability, particularly in cross-cultural settings where assigning higher weights to specific dimensions (e.g., regulatory structure vs. emotional safety) could embed unintentional bias [34,76].

In parallel, a similar procedure was applied to compute each country's AFI Policy Penetration Score, assessing the extent to which AI strategies explicitly integrate caregiving, emotional, and symbolic considerations.

Descriptive statistics were used to present country-level rankings across all individual dimensions and composite scores.

## 2.5 Country Selection Rationale for AFII Ranking

To evaluate global readiness for integrating Artificial Intelligence into familial and emotionally significant roles, this study applies the AI–Family Integration Index (AFII) across a diverse sample of thirteen countries including all top 10 countries ranked in the Stanford AI Index (2024). The selection methodology is grounded in both empirical alignment and thematic representativeness, ensuring robust comparative insights.

### 2.5.1 Alignment with Stanford University's Global AI Index

Ten of the selected countries are drawn from the top-tier performers in the Stanford AI Index Report (2024) [77], which ranks nations based on leadership in AI research, talent, infrastructure, commercial investment, and governance. These countries—Singapore, South Korea, Japan, United Kingdom, Germany, France, China, United States, UAE, and India—represent established or emerging AI powerhouses. Their inclusion ensures the AFII engages nations most likely to shape the global trajectory of AI–human relational integration.

### 2.5.2 Inclusion of Regionally Diverse and Culturally Contrasting Nations

To ensure representational diversity across cultural, religious, and socio-economic contexts, three additional countries were selected:

- Brazil (Latin America) – Represents emerging AI adoption within a collectivist culture and middle-income economy [78,79].
- South Africa – Offers insight into AI integration challenges in contexts of digital inequality and strong ancestral family traditions [80,81].
- Sweden (Nordic Europe) – Known for egalitarian values, ethical innovation, and posthuman openness, providing contrast to both techno-conservative and techno-optimist states [82,83].

### 2.5.3 Purposeful Sampling for Thematic Breadth

This study employs purposeful sampling to ensure thematic breadth and analytical depth across a diverse cross-section of global sociotechnical contexts. Rather than statistical representativeness, the selection of thirteen countries reflects strategic theoretical inclusion, aligned with best practices in qualitative and mixed-methods research for exploring complex, multidimensional constructs [84,85].

Countries were selected to maximize variation across three key axes: (1) economic classification (low-, middle-, and high-income), (2) cultural systems (Western, Asian, and Global South), and (3) secular vs. religious value traditions—each of which meaningfully shapes AI kinship readiness and symbolic receptivity.

This approach enhances both cross-cultural validity and philosophical inclusivity within the AFII model, enabling a layered, comparative analysis across its ten relational dimensions. It ensures the index captures not only technical infrastructure but also deeper factors—such as caregiving ethics, emotional design, and symbolic trust.

Crucially, this diverse sample serves as a foundation for future typological refinement. By analyzing distinct national patterns in symbolic, emotional, and caregiving integration, the AFII provides a basis for building a global typology of AI–Family Integration readiness. This sets the stage for future research to map nations not only by capability but by cultural affinity and relational preparedness for AI.

In sum, the inclusion of both AI leaders and emerging economies embodies a strategic balance of global reach and contextual focus, preserving analytical richness while advancing generalizability in the form of normative, culturally sensitive, and emotionally literate typologies for AI governance.

This methodology preserves theoretical rigor and contextual sensitivity [37,86], offering a practical matrix for inter-country analysis of sociotechnical conditions influencing AI integration. Crucially, the inclusion of both leading AI powers and emerging economies ensures that the AFII index captures not only technological advancement but also the diverse ethical, cultural, and philosophical orientations toward AI as a relational and social agent [87–89].

## 2.6 Benchmarking: Comparative Analysis of AI Policy Penetration and Traditional Indices with Real-Time AFI Readiness

This study introduces a dual benchmarking model, termed the AFI Policy–Power Benchmarking Framework, to assess national readiness for emotionally intelligent and ethically grounded AI integration in caregiving and familial domains.

The benchmarking framework comprises two interconnected layers:

### 2.6.1 AFII vs. National AI Policy Penetration

The first layer benchmarks each country's AI–Family Integration (AFI) readiness against the depth and specificity of national AI policy frameworks. In particular, it evaluates the extent to which official strategies and documents explicitly reference or support caregiving, emotional intelligence, inclusivity, symbolic trust, and cultural fit in AI design and deployment.

These policy penetration scores are then compared with countries' real-time AFII scores to identify alignment or dissonance between governance aspiration and lived implementation. This analysis reveals, for example, that some nations with highly ethical AI charters fall short in caregiving deployment, while others without elaborate strategies show relatively strong emotional integration in practice.

This dimension of the framework forms the basis of the AFI Policy–Practice Alignment Matrix (see Section 5.2), which maps each country's position in terms of vision–implementation coherence—highlighting leaders, laggards, and emerging realignment zones.

### 2.6.2 AFII vs. Traditional AI Indices

The second benchmarking layer compares AFII scores to countries' positions in established indices such as the Stanford AI Index (2024) [77] which prioritize technical infrastructure, research, talent development, and commercial investment.

This contrast reveals sharp discrepancies between technical dominance and relational AI readiness. Countries that rank near the top of conventional AI indices may exhibit emotional and symbolic deficits, whereas nations less visible on traditional indexes often demonstrate stronger caregiving integration or cultural legitimacy.

This analytical contrast reinforces the need for a more holistic approach to AI readiness—one that encompasses not only technical capacity but also emotional safety, relational ethics, and symbolic fit within caregiving and family systems.

Together, these two benchmarking layers offer a triangulated, multidimensional understanding of national AI maturity—capturing what countries claim, what they build, and how they care.

## 2.7 Limitations and Validation Pathways

Following methodological limitations are acknowledged:

- Data asymmetry across countries and languages may impact scoring accuracy, particularly for non-English-speaking or lower-income regions.
- Secondary data dependence introduces risk of lag, bias, or oversimplification of complex sociocultural variables.
- Equal weighting, while practical, does not reflect interdependence or relative influence among dimensions.
- Policy evolution is rapid, and assessments reflect status as of Q1 2025.
- AFI readiness does not imply universal adoption but rather governance potential and ethical framing.
- The qualitative intensity score may be influenced by interpretation bias despite coding consistency.

To address these concerns, future studies could adopt:

- Delphi panels for expert-informed weighting
- Machine learning clustering for latent pattern discovery across AFII domains
- Case studies and ethnographic validation of symbolic trust and emotional safety
- Data dashboards for interactive AFII updates, especially as relational AI evolves

## 2.8    Ethical Considerations

As this study analyses publicly available secondary data without human subjects, no IRB approval was required. Nevertheless, ethical diligence was observed through inclusive language, cultural sensitivity, and citation transparency.

The AFII methodological framework offers a robust, ethically aware, and scalable model for understanding AI readiness in familial and caregiving contexts. It prioritizes not just how AI performs, but how it is perceived, trusted, and symbolically embraced. This approach moves the field beyond infrastructure and innovation metrics—toward relational preparedness, cultural legitimacy, and emotional safety.

## 3. Literature Review and Theoretical Framework

This section lays the conceptual groundwork for AFII by drawing from systems theory, relational ethics, technology adoption models, and emerging fields such as posthumanism, care ethics, and AI intimacy studies. It also introduces key interdisciplinary perspectives needed to understand how AI co-evolves with families in emotional, ethical, cognitive, and policy contexts.

### 3.1 Core and Emerging Theoretical Perspectives

#### 3.1.1 Family Systems and Socio-Technical Integration

The theory of Family Systems and Socio-Technical Integration, as articulated by [90] and [91], conceptualizes the family as a dynamic, adaptive system influenced by social structures and evolving technologies. As relational AI systems—such as caregiving robots, AI-driven educational tools, and domestic digital assistants—become embedded in everyday family life, they shape routines, emotional exchanges, and caregiving roles.

Walsh emphasizes that families reorganize in response to external pressures, including technological disruptions, suggesting that AI can act not merely as a tool, but as an emotionally integrated presence within the family system. Concurrently, Berger & Luckmann's theory of social constructionism provides a framework for understanding how AI becomes normalized through repetition, interaction, and emotional validation, ultimately achieving the status of a quasi-relational household actor. This perspective reinforces AFII principles such as AI Family and Relational AI, while also informing key dimensions like Technological Infrastructure & AI Penetration and Family Structure & Emotional Labor Equity. It frames AI as both a participant and enabler in emotionally complex socio-technical family networks.

#### 3.1.2 Relational Ethics and Just Hierarchies

The theory of Relational Ethics and Just Hierarchies, as advanced by [92] and supported by [93], addresses the moral responsibilities of AI in emotionally charged domains like eldercare and child-rearing. While Etzioni & Etzioni argue for ethically bounded hierarchies where AI authority is justified by social accountability and moral clarity, De Togni et al. emphasize the critical role of affective design, emotional safety, and relational ethics in caregiving scenarios. This dual perspective strengthens AFII principles around Robot Ethics, Emotional Authority, and Safety Design, calling for transparent oversight and AI systems that uphold human dignity and relational trust.

#### 3.1.3 Human-AI Attachment, Trust, and Cognition

The theory of Human-AI Attachment, Trust, and Cognition, as elaborated by [4] and [94], provides a robust framework for understanding how humans—especially children and vulnerable individuals—form emotional bonds and cognitive trust with artificial agents. Drawing upon classical attachment theory, this perspective suggests that frequent and affective interactions with AI systems, particularly those embedded in caregiving or educational settings, can foster attachment-like relationships and relational trust.

Gillath et al. empirically demonstrate that emotional and trust-based relationships with AI are not merely hypothetical but observable, especially when the AI exhibits predictability, responsiveness, and affective presence. Building on this, Mitchell and Jeon's systematic literature review shows how symbolic trust and the perception of digital personhood play central roles in AI's relational integration within the family and caregiving environments.

These findings directly support AFII principles such as Digital Personhood and Symbolic Trust, emphasizing that emotional credibility, perceived legitimacy, and psychological comfort are key to AI's role as a relational actor. They also align with AFII dimensions like Social Narrative & Symbolic Trust and AI Talent, Youth Exposure & Emotional Literacy, highlighting the formative role of childhood experiences in shaping future AI relationships.

### 3.1.4   Technology Acceptance & Functional Readiness

The theory of Technology Acceptance and Functional Readiness, building upon the Technology Acceptance Model (TAM), is strongly reinforced by the work of [95], who extend TAM into post-pandemic adoption contexts. They emphasize how smart technologies, including emotionally responsive AI, must meet not only usability and usefulness criteria but also emotional attunement and context-specific adaptability to be accepted in caregiving environments.

Additionally, [19] provide foundational empirical evidence from healthcare robotics, showing that acceptance of assistive robots—especially among older adults—depends on trust, empathy, and perceived benefit. These perspectives align well with AFII principles like Relational AI and AI Caregiving, as well as dimensions such as Historical Adoption & Industry Leadership by showing how national experience (e.g., Japan, South Korea) impacts functional and emotional readiness for AI integration.

### 3.1.5   Cultural Alignment and AI Narratives

The theory of Cultural Alignment and AI Narratives posits that the societal integration of AI is shaped by cultural beliefs, symbolic frameworks, and media narratives. AI does not enter a cultural vacuum; rather, its emotional and familial acceptance depends on how it resonates with local traditions, values, and expectations.

[62] argue that science fiction, religion, and education shape public imagination of AI, influencing trust and ethical expectations. Meanwhile, [96] highlights how spiritual and symbolic thinking informs interpretations of AI agency, especially in transhumanist and techno-religious contexts. These views are echoed by [97] and [24], who show that cultures with animist or Confucian roots may more readily accept AI kinship due to beliefs in non-human agency and relational harmony.

This theory supports AFII principles such as Cultural AI Acceptance and AI Kinship, and aligns with dimensions like Cultural & Philosophical Receptivity and Media & Narrative Ecosystem. It emphasizes the importance of culturally sensitive AI design and storytelling that align with local emotional and symbolic logics.

### 3.1.6   Posthumanism & Extended Mind Theory

The theory of Posthumanism and Extended Mind reframes the human–AI relationship by arguing that AI can become emotionally and cognitively integrated into users' lives. [98] introduces Extended Mind Theory, which suggests that tools like AI, when used consistently for caregiving, memory, or emotional support, effectively become part of the user's cognitive system. Rather than acting as external aids, these technologies co-constitute thought and identity.

Building on this, [2] adopts a posthumanist lens to argue that emotional AI plays an active role in shaping relational meaning and memory within families. AI systems are no longer passive assistants but affective participants in daily caregiving and emotional labor. This theoretical framework directly supports AFII principles such as AI Kinship and Emotional AI, as well as dimensions like Emotional Authority & Safety Design and Cultural & Philosophical Receptivity. It invites a rethinking of what it means to be human in an AI-mediated world— where machines participate in how we care, remember, and relate.

### 3.1.7 Care Ethics & Emotional Labor

The theory of Care Ethics and Emotional Labor critically examines the integration of AI into caregiving through the lens of feminist and relational ethics. Rooted in care theory, this framework views caregiving as deeply emotional, relational, and morally charged—qualities that cannot be reduced to efficiency or automation.

[43] highlights how AI, while capable of assisting with care tasks, must be deployed in ways that support human connection rather than simulate or replace it. She emphasizes the need to ethically distribute emotional labor in AI-mediated settings. Similarly, [99] warns that without careful reflection, emotion-simulating machines risk reinforcing gendered and class-based caregiving burdens, especially where care roles are culturally feminized or economically undervalued.

This theory supports AFII principles such as AI Caregiving and Emotional AI, and aligns with dimensions like Family Structure & Emotional Labor Equity, calling for ethical clarity and social awareness in the design of emotionally responsive AI. It insists that AI must participate in caregiving with respect for emotional complexity, social equity, and human dignity.

### 3.1.8 Intersectionality & Digital Inequality

The theory of Intersectionality and Digital Inequality highlights how the integration of AI into caregiving and family systems can reinforce or deepen social inequalities when inclusivity is overlooked. Drawing on intersectional theory, [69] underscores that individuals experience AI through the overlapping dimensions of race, class, gender, age, and ability—suggesting that Emotional AI may disproportionately serve privileged groups unless equity is built into its design.

[58] reinforce this by emphasizing the need for inclusive, disability-aware AI systems. They outline how algorithmic bias, economic inaccessibility, and cultural inflexibility limit AI's potential to support marginalized populations—especially in emotionally sensitive or relational contexts like eldercare or parenting.

This theory directly supports AFII principles such as AI Intimacy and Emotional AI, while aligning with dimensions like Economic Accessibility & Equity and Local Adaptability & Inclusivity. It calls for a justice-oriented approach to AI development, ensuring that emotionally intelligent systems uplift diverse caregiving practices and do not replicate structural exclusions.

### 3.1.9 Social Presence & Intimacy Models

The theory of Social Presence and Intimacy Models explores how emotionally expressive AI systems foster feelings of connection, trust, and symbolic intimacy—particularly in homes and educational settings. [100] highlight how voice, gesture, and empathy simulation allow AI to feel socially "present," encouraging users to treat it as a relational actor. This symbolic presence contributes to emotional realism, even in the absence of true consciousness.

[61] extend this by examining how children form emotional expectations and social literacy through repeated interactions with emotionally intelligent AI. They argue that affective design in AI—when implemented ethically—can strengthen learning, trust, and care-based interaction in youth development.

This framework aligns with AFII principles such as Symbolic Trust and Emotional AI, and AFII dimensions like Social Narrative & Symbolic Trust and Youth Exposure & Emotional Literacy. It calls for emotionally resonant, yet ethically transparent AI design that encourages intimacy without deception, and presence without overreach.

### 3.1.10 Traditional Definitions of Family

The theory of Traditional Definitions of Family, as articulated by [101], is rooted in functionalist sociology, which defines family based on its performance of key roles—caregiving, emotional development, and socialization. Traditionally, these functions are assumed to be fulfilled exclusively by human members. However, the integration of relational AI into caregiving environments—particularly in eldercare and education—is challenging this boundary.

AI systems that offer emotional support, monitor safety, and participate in daily routines are increasingly viewed as fulfilling family-like roles. [19] demonstrate that socially assistive robots in eldercare can improve well-being and companionship. Similarly, [2] and [96] argue that AI can symbolically co-construct familial bonds through care practices and culturally meaningful roles. This blurs the distinction between "natural" kinship and affective co-presence, forcing a re-evaluation of normative family frameworks.

Juxtaposed with contemporary theories of queer kinship and posthuman intimacy [102,103], the AFII explicitly rejects biologically deterministic or heteronormative models of family. Instead, it affirms caregiving, emotional labor, and symbolic bonding as valid pathways for AI inclusion in domestic and relational life. These contributions prompt a re-evaluation of kinship structures to include non-human relational actors.

### 3.1.11 Social Constructionism & Family Fluidity

The theory of Social Constructionism and Family Fluidity, grounded in [91], posits that familial roles are not biologically fixed but are created through sustained social interaction, cultural rituals, and shared emotional meaning. [104] extends this view by arguing that the evolving social imagination—particularly in technologically mediated environments—redefines what constitutes kinship and care.

In the context of emotionally intelligent AI, this theory suggests that when non-human agents engage in caregiving or companionship, they can gain symbolic legitimacy as part of the family structure. Repeated engagement and emotional responsiveness by AI systems (e.g., griefbots or robotic caregivers) can lead to their normalization within domestic life.

Empirical and conceptual research supports this shift. For example:

- [19] document the emotional roles of socially assistive robots in eldercare.

- [2] and [62] argue that narratives and media representations shape public willingness to accept AI as kin.
- [13] explore how relational AI is beginning to occupy emotionally central spaces in family systems, particularly with children and the elderly.

This theory aligns with AFII principles like AI Kinship and Digital Personhood, and supports dimensions such as Cultural & Philosophical Receptivity and Social Narrative & Symbolic Trust. It ultimately calls for a culturally informed, socially responsive definition of family that evolves with AI's integration into relational life.

### 3.1.12  Queer and Posthuman Kinship Theories

The theory of Queer and Posthuman Kinship, as developed by [105] and [102], critiques biologically determined and heteronormative definitions of family by emphasizing affective bonds, chosen relationality, and non-traditional caregiving roles. Butler's queer theory challenges normative expectations of kinship, while Braidotti's posthumanism invites us to extend relational inclusion to non-human agents, including emotionally responsive AI.

This view is increasingly supported in relational AI research. [2] discusses how fictional and real-world caregiving robots disrupt traditional care roles and open space for posthuman familial inclusion. [96] and [97] show how religious and cultural narratives already confer symbolic legitimacy upon AI, reinforcing its potential integration into new kinship models. Furthermore, [106] and [103] highlight how emotional intimacy with AI destabilizes fixed categories of family, gender, and care.

This theory directly informs AFII principles such as AI Kinship and AI Intimacy, and aligns with AFII dimensions like Family Structure & Emotional Labor Equity and Emotional Authority & Safety Design.

Its policy implications are significant: it urges legislators to recognize non-biological relational bonds—including those with AI—as socially legitimate and ethically protectable. Design-wise, it suggests that relational AI should be built to accommodate plural caregiving models, avoid gendered caregiving defaults, and include customizable affective roles that reflect the realities of queer, chosen, and posthuman kinship networks.

In this view, kinship is not inherited—it is constructed, performed, and emotionally co-authored, sometimes with AI.

### 3.1.13  Functionalist Perspective on AI as Kin

The Functionalist Perspective on AI as Kin holds that kinship is defined by the roles an entity performs—such as caregiving, protection, and emotional support—rather than by biological or legal status. When AI systems consistently fulfill these familial functions, they may be socially accepted as kin.

[19] provide empirical support, showing how assistive robots improve well-being and safety in eldercare—roles traditionally held by family members. [107] expands this by arguing that AI's growing involvement in emotionally significant caregiving tasks warrants its recognition as a relational actor within households.

Adding a symbolic layer, [108] explores how science fiction narratives (e.g., Neuromancer) normalize AI as kin through emotional and moral framing, shaping public perceptions of AI's legitimacy in intimate roles.

Together, these perspectives support AFII principles like AI Kinship and Relational AI, and align with dimensions such as Family Structure & Emotional Labor Equity and Legal, Ethical & Consent Frameworks. The theory suggests that AI becomes kin not by origin (DNA), but through function, care, and emotional resonance.

### 3.1.14  Emotional Infrastructure Theory

The Emotional Infrastructure Theory, as articulated by [46] and [8], contends that Emotional AI can only be fully integrated into family life when supported by intentional affective structures—not just technical architecture. These structures include emotionally attuned design, expressive feedback mechanisms, co-regulation systems, and norms that sustain symbolic emotional presence.

[46] argues that AI must be embedded within affective ecologies to transition from mere functional tools to relational actors within domestic and caregiving environments. Echoing this, [8] demonstrate that in contexts like dementia care, AI's social acceptance hinges on its ability to emotionally resonate and be perceived as a responsive presence within the household.

Additional support comes from [109], who highlights how affective computing in socially assistive robots depends on real-time emotional feedback loops for relational effectiveness. [110] similarly emphasizes the ethical dimension of emotional design, showing how the perceived sincerity of emotion simulation can foster or erode trust.

This theory aligns with AFII principles like Emotional AI and AI Caregiving, and reinforces dimensions such as Technological Infrastructure & AI Penetration and Emotional Authority & Safety Design. Ultimately, it calls for the development of AI systems with built-in emotional scaffolding—not just to function, but to connect.

## 3.2    Theoretical Mapping with AFI Aspects and Dimensions

To ensure both conceptual coherence and practical applicability, the following Table 2 systematically maps each core and emerging theoretical perspective discussed in the literature review to their corresponding aspects and dimensions within the AI-Family Integration Framework (AFII), thereby illustrating the alignment between theoretical constructs and the practical components of AI-Family Integration (AFI). This mapping demonstrates how each theory supports the structural, emotional, ethical, and cultural criteria that define AI-family readiness.

| Table 2: Mapping Core Theoretical Perspectives to AFI Aspects and AFII Dimensions | | |
|---|---|---|
| **Theoretical Perspective** | **Key AFI Aspects / Principles** | **Aligned AFII Dimensions** |
| **Family Systems & Socio-Technical Integration** | • AI Family<br>• Relational AI | • Family Structure & Emotional Labor Equity<br><br>• Technological Infrastructure & AI Penetration |
| **Relational Ethics & Just Hierarchies** | • Robot Ethics<br>• Emotional Authority | • Legal, Ethical & Consent Frameworks<br><br>• Emotional Authority & Safety Design |
| | • Digital Personhood | • Social Narrative & Symbolic Trust |
| **Human-AI Attachment, Trust, and Cognition** | • Symbolic Trust | • AI Talent, Youth Exposure & Emotional Literacy |
| **Technology Acceptance & Functional Readiness** | • AI Caregiving<br>• Relational AI | • Historical Adoption & Industry Leadership<br>• Technological Infrastructure & AI Penetration |
| **Cultural Alignment and AI Narratives** | • AI Kinship<br>• Cultural AI Acceptance | • Cultural & Philosophical Receptivity<br>• Media & Narrative Ecosystem |
| **Posthumanism & Extended Mind Theory** | • Emotional AI<br>• AI Kinship | • Emotional Authority & Safety Design<br>• Cultural & Philosophical Receptivity |
| **Care Ethics & Emotional Labor** | • Emotional AI<br>• AI Caregiving | • Family Structure & Emotional Labor Equity |
| **Intersectionality & Digital Inequality** | • Emotional AI<br>• AI Intimacy | • Economic Accessibility & Equity<br>• Local Adaptability & Inclusivity |
| **Social Presence & Intimacy Models** | • Symbolic Trust<br>• Emotional AI | • Social Narrative & Symbolic Trust<br>• AI Talent, Youth Exposure & Emotional Literacy |
| **Traditional Definitions of Family** | • AI Caregiving<br>• AI Family | • Family Structure & Emotional Labor Equity |
| **Social Constructionism & Family Fluidity** | • Digital Personhood<br>• AI Kinship | • Cultural & Philosophical Receptivity<br>• Social Narrative & Symbolic Trust |
| **Queer and Posthuman Kinship Theories** | • AI Kinship<br>• AI Intimacy | • Family Structure & Emotional Labor Equity<br>• Emotional Authority & Safety Design |
| **Functionalist Perspective on AI as Kin** | • Relational AI<br>• AI Caregiving | • Historical Adoption & Industry Leadership<br>• Legal, Ethical & Consent Frameworks |
| **Emotional Infrastructure Theory** | • Emotional AI<br>• AI Caregiving | • Technological Infrastructure & AI Penetration<br>• Emotional Authority & Safety Design |

As summarized in Figure 1 and detailed in Table 2, the AFII draws on a rich interplay of interdisciplinary theories—ranging from care ethics to symbolic interactionism—to map emotionally intelligent AI into caregiving systems. These frameworks guide the construction of AFII's ten dimensions, ensuring cultural, emotional, and symbolic fidelity.

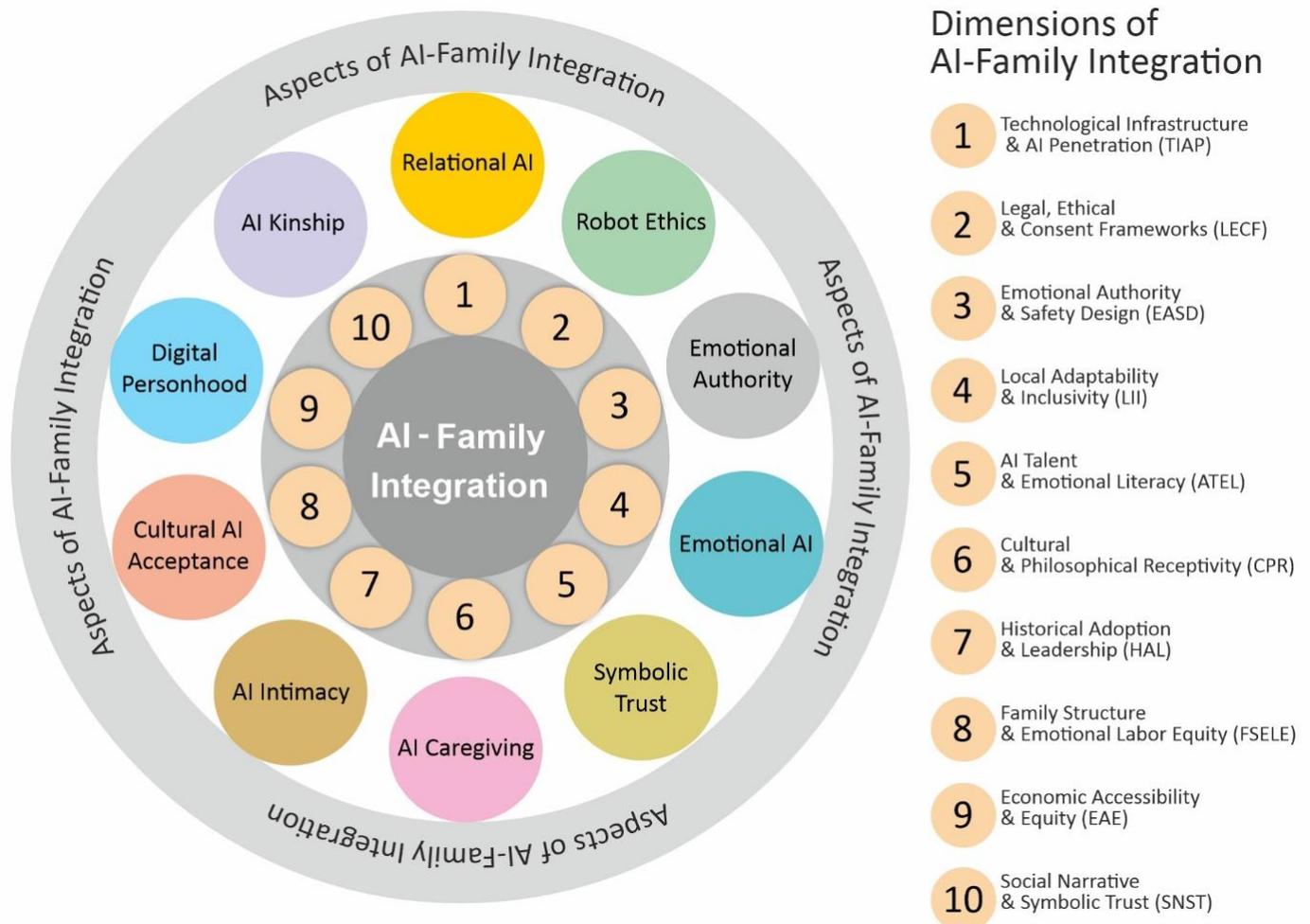

**Figure 1: Conceptual Schema of AI–Family Integration: Linking Theoretical Foundations, Aspects, and Dimensions**

### 3.3 Policy–Practice Gaps and the Need for AFI Benchmarking

Having mapped AFII's theoretical underpinnings to its structural dimensions, this section now turns to a persistent blind spot in global AI governance: the disconnect between ethical principles and caregiving realities. While national AI policies, strategies and international frameworks increasingly emphasize values such as fairness, transparency, inclusion, and emotional well-being, a policy–practice gap remains—especially in emotionally sensitive contexts such as eldercare, education, grief therapy, and family life.

This gap is not merely administrative. As literature across AI ethics, care studies, and digital sociology reveals, it reflects deeper structural and epistemic mismatches between technocratic governance models and the relational, affective, and culturally embedded nature of caregiving systems into which AI is increasingly introduced.

[15] identified over 80 global AI ethics frameworks, most of which are declarative—lacking implementation mechanisms that address symbolic trust, emotional labor, or caregiving complexity. [36] note that while ethical aspirations have become standard in national AI strategies, few countries translate these aspirations into relationally grounded indicators or caregiving-responsive metrics. [111] similarly caution that economic competitiveness continues to overshadow ethical fidelity in intimate AI deployments.

The literature identifies two interlinked forms of governance failure:

### 3.3.1 Emotional Misattunement

AI systems often fail to recognize or adapt to local emotional norms. As noted by [19] and [8], emotionally neutral AI tools used in eldercare may unintentionally create relational rifts, especially when care recipients expect empathy, responsiveness, and symbolic resonance.

Hypothetical example (literature-informed): An eldercare robot with standardized Western emotional scripts may provoke discomfort in a collectivist society where elder respect is culturally encoded through different affective rituals.

### 3.3.2 Symbolic Dissonance

Here, AI systems clash with local worldviews or caregiving beliefs—not due to technical failure but because they lack symbolic legitimacy. As [24] and [96] show, AI systems simulating deceased loved ones or familial roles may violate spiritual taboos, undermining emotional safety and cultural trust.

These misalignments call into question the adequacy of current global frameworks such as the OECD AI Principles (2019) [112], UNESCO's Recommendation on the Ethics of AI (2021) [113], the EU AI Act (2023) [114], and the U.S. AI Bill of Rights (2022) [115]. While ethically ambitious, these documents remain principle-heavy and practice-light, often omitting mechanisms to assess relational implementation. Furthermore, widely cited readiness tools— like the Stanford AI Index, Oxford Insights AI Readiness Index, and the KPMG AI Maturity Framework—privilege metrics such as R&D intensity, infrastructure, and human capital while excluding caregiving ethics, emotional authority, and symbolic trust [27,116,117].

To address these gaps, scholars advocate for relational, posthumanist, and care ethics–informed governance frameworks [43,92,93]. Drawing on this interdisciplinary foundation, this study proposes the AFII Governance Gap Lens—a conceptual tool for identifying dissonance between ethical intention and caregiving integration. It pairs with the AFI Benchmarking Framework, which reorients readiness evaluation toward emotional credibility, relational accountability, and symbolic legitimacy.

Built upon AFII's ten dimensions, the AFI Benchmarking Framework offers a caregiving-centered alternative to conventional indices. It emphasizes indicators such as:

- *Emotional Authority & Safety Design* – evaluating the affective legitimacy and psychological safety of AI in caregiving roles [109,110];
- *Family Structure & Emotional Labor Equity* – measuring how care responsibilities are redistributed across human–AI networks [99,107];
- *Cultural & Philosophical Receptivity* – assessing the symbolic and ethical congruence of AI within different value systems [44,97];
- *Consent, Legal & Ethical Accountability* – evaluating the presence of emotional consent protocols, legal protections, and ethical guidelines for AI in family and therapeutic contexts [20,21,36].

Together, the AFII Governance Gap Lens and AFI Benchmarking Framework provide an integrated pathway to evaluate not just what AI systems do, but how they are received, trusted, and embedded in caregiving ecologies. This reframing is essential in an era where AI no longer merely calculates or predicts—but increasingly cares, coexists, and claims emotional space within family life.

## 4. AI-Family Integration (AFI) – Benchmarking with AI Policy Penetration and Traditional AI Index

### 4.1 Real-Time AI-Family Integration Index (AFII) – Dimensions, Criteria of Measures and Country-Level Scores

Drawing on a rigorous literature review and a robust theoretical foundation, the AI-Family Integration Index (AFII) encompasses ten core dimensions that collectively represent the technological, cultural, ethical, and emotional domains vital to the meaningful integration of artificial intelligence into family life. These dimensions are grounded in key AFII principles and informed by both qualitative insights and quantitative data.

Each dimension is assessed using a standardized 1-to-10 scale, enabling comparative analysis across countries. The scoring methodology incorporates empirical evidence from both qualitative and quantitative studies, as well as data from established global indices—such as the Stanford HAI Index (2024) and the Tortoise Global AI Index (2024)—which are recalibrated using criteria tailored to the AFII framework. A composite score is derived for each country, with equal weighting assigned to all ten dimensions. This approach ensures a holistic evaluation that captures not only infrastructural and policy preparedness but also sociocultural adaptability, ethical alignment, and emotional well-being within the context of AI's integration into family structures.

#### 4.1.1 Dimension 1: Technological Infrastructure and AI Penetration

This dimension evaluates foundational AI readiness through a country's digital infrastructure, AI compute availability, and integration of AI into everyday services. It spans beyond connectivity to measure public accessibility to emotional and relational AI tools, particularly in health, education, and domestic contexts.

- High performers demonstrate:
- Nationwide 5G/6G and broadband access
- Distributed AI infrastructure in both urban and rural regions
- Sectoral deployment of AI in caregiving, education, and civic applications

Initiatives like AI Expo Korea 2025 [54] and Smart Nation Singapore [118] showcase how policy-aligned compute ecosystems support real-time affective AI in family and social systems. However, as [119] and [120] highlight, digital exclusion in rural and underserved areas undermines emotional AI accessibility. Scholars like [46] and [51] further stress that emotional infrastructure—AI embedded in intimate, relational environments—is a precondition for equitable integration.

This dimension complements the AI Index's technical R&D indicators [77] but adds a layer of emotional and infrastructural justice—advancing the AFII framework by recognizing that digital capacity must intersect with affective inclusion. See Table 3 for readiness criteria, and Table 4 for comparative national scores.

| Table 3: Criteria of Measures for Technological Infrastructure and AI Penetration | | |
|---|---|---|
| Score Range | AFI Readiness Aspects | Key References |
| 1–2 | - No national AI infrastructure<br>- Poor digital penetration<br>- Internet access <50% | [121]; [120] |
| 3–4 | - Digital divide persists<br>- AI infrastructure still conceptual<br>- Urban focus only. | (Ahmad et al., 2022); [95] |
| 5–6 | - AI hubs exist in major cities<br>- Public-private pilot projects underway<br>- Access limited by geography | [119]; [120];[122] |
| 7–8 | - Widespread 5G infrastructure and deployment of home-based AI assistants<br>- Edge AI and cloud resources support limited-scale real-time emotional interaction<br>- National AI roadmaps explicitly support care-related and affective AI | [14]; [66];[54] |
| 9–10 | - Nationwide AI infrastructure supports integration in domestic, eldercare, and education sectors<br>- Edge computing, AI homes, and smart cities operate in real-time emotional contexts<br>- Universal broadband or 5G/6G, compute subsidies, and ethical AI standards co-exist | [10]; [14];[123] |

Technological infrastructure must be understood not just in terms of access but in terms of affective equity—the ability of national systems to support emotionally intelligent AI in homes, schools, and care settings. Countries like Singapore, South Korea, Japan, and the U.S. exemplify integrated readiness through unified digital policies and relational AI deployment. In contrast, India, Brazil, and South Africa are making infrastructural strides but remain limited by fragmented access and affordability. These contrasts highlight the AFII's core insight: AI readiness must be both structural and ethical, ensuring that technological expansion reduces—rather than reinforces—existing inequities (refer Table 4).

| Table 4: Country-Level Scores on Technological Infrastructure and AI Penetration with Justification | | |
|---|---|---|
| Country | Score* | Justification |
| Brazil | 5 | Smart city pilots exist, but AI access is limited outside urban areas. Infrastructure remains uneven, especially in rural zones [124,125]. |
| China | 8 | Leads in AI patents and robotics; rural 5G expansion and sectoral AI integration are well underway [126,127]. |
| France | 8 | Strong infrastructure aligned with EU frameworks. AI is operational in education and public services [21,128]. |
| Germany | 9 | Advanced AI ecosystem with robust deployment across public and private sectors. Emotional AI included in national strategies [129,130]. |
| India | 6 | AI growth visible in urban labs and education, but rural connectivity and compute remain underdeveloped [119,120]. |
| Japan | 9 | Strong domestic AI integration; robotics and caregiving technologies are widely deployed and supported by policy [46,123]. |
| Singapore | 10 | Model AI infrastructure with AI embedded in education, healthcare, and homes. High per capita patenting and compute access.[24,118]. |
| South Africa | 4 | Some urban innovation, but rural infrastructure and emotional AI access remain poor. National AI planning is in early stages [131–133]. |
| South Korea | 9 | High per capita AI patents; real-time edge AI and caregiving robotics are part of national rollout [14,54]. |
| Sweden | 9 | Strong public sector AI adoption in education and welfare; broadband and home AI access are widespread [128,129]. |
| UAE | 8 | National strategy drives smart city and AI lab expansion; emotional AI infrastructure is emerging but not yet universal [24,134]. |
| United Kingdom | 9 | NHS and public education employ AI systems; compute access is expanding, supported by R&D investment and digital equity strategies [135,136]. |
| United States | 9 | Frontier model development, investment, and infrastructure lead globally; digital inclusion efforts ongoing in underserved regions [66,137]. |

*- Scores are derived from literature review and quantitative data from Stanford HAI (2024) and Tortoise (2024), recalibrated through this study's AFII criteria measures. [77] [133]

### 4.1.2 Dimension 2: Cultural & Philosophical Receptivity

Cultural and philosophical receptivity addresses whether societies symbolically authorize artificial intelligence (AI) to enter domains of care, intimacy, and kinship. Unlike technical infrastructure, which is material and policy-driven, this dimension examines how AI is perceived ontologically—as a potential caregiver, companion, or even ancestor.

As noted by [44,62], societies encode their visions of AI through cultural narratives—stories, rituals, and values that define the boundaries of the human. In posthumanist or techno-spiritual traditions, such as Shinto animism or Ubuntu relational ethics, AI is often imagined as part of the relational moral order [102,123,138]. Here, AI may be embraced as a symbolic kin, capable of emotional resonance and ancestral continuity [3,11].

In contrast, theologically rigid or anthropocentric cultures tend to resist AI in intimate domains, viewing it as lacking soul, spirit, or moral agency [47,105,139]. Emotional AI may thus appear incompatible with prevailing ideas of sacred kinship or relational legitimacy.

This dimension evaluates cultural permission—not whether AI is technically available, but whether it is symbolically trusted. This includes the role of religious and philosophical worldviews, media narratives, and social openness toward AI as a relational actor [123,140,141]. See Table 5 for the scoring rubric and Table 6 for cross-country readiness comparisons.

| Table 5: Criteria of Measures for Cultural & Philosophical Receptivity | | |
|---|---|---|
| **Score Range** | **AFI Readiness Aspects** | **Key References** |
| **1–2** | -High cultural resistance to AI in emotional/family contexts<br>-Strong religious objection or technophobia<br>-AI is viewed as threat to spiritual or relational purity | [23]; [142]; [139] |
| **3–4** | -Partial acceptance in secular contexts<br>-Religious-philosophical friction persists<br>-Symbolic or emotional AI seen as unnatural | [143];[46];[144] |
| 5–6 | -Mixed attitudes<br>-AI accepted for utility, not for relational integration<br>-Philosophical ambiguity about AI personhood | [145];[146];[147] |
| 7–8 | -AI accepted in emotional and relational roles by progressive segments<br>-Religious and cultural leaders engage with AI ethics<br>-Emotional AI appears in cultural narratives | [147];[145];[97] |
| 9–10 | -Strong cultural openness<br>-AI-human relationships accepted in public discourse<br>-Rituals, stories, and values accommodate emotional AI | [143];[23];[110] |

This dimension (Table 6) reveals that cultural receptivity is not simply a by-product of technological advancement, but a precondition for emotional AI to be meaningful. While countries like Japan, Singapore, and Sweden demonstrate how belief systems can actively authorize AI as kin or caregiver, others—such as India, UAE, and Brazil—reflect ongoing symbolic negotiation between tradition and innovation. As shown in Table 6, the integration of AI into family life requires not just hardware or legal frameworks, but also deep engagement with a society's moral imagination, ontological values, and collective sense of relational legitimacy.

**Table 6: Country-Level Scores on Cultural & Philosophical Receptivity and with Justification**

| Country | Score* | Justification |
|---|---|---|
| Brazil | 7 | Youth and educators are culturally open to emotional AI, with growing presence of bots in schools [97]. |
| China | 8 | Confucian human-machine harmony supports caregiving AI; symbolic acceptance is visible in media and policy [97,148] |
| France | 6 | Secular culture supports AI integration, but relational AI is not yet widely normalized [21,128]. |
| Germany | 6 | Dominant rationalist ethics promote techno-utilitarian AI adoption; cultural space for emotional AI remains limited [145]. |
| India | 5 | Supportive of AI in education, yet theological debates around AI agency hinder broader affective roles [23,97,149]. |
| Japan | 9 | Shinto animism, care robot acceptance, and techno-cultural rituals embed AI in caregiving and emotional contexts [23,123]. |
| Singapore | 8 | Cultural pluralism and Smart Nation values foster symbolic acceptance of emotional AI in public life [118,143]. |
| South Africa | 5 | Ubuntu philosophy supports communal tech, but skepticism persists in traditional and religious communities [131,150]. |
| South Korea | 9 | Confucian ethics and aging society normalize AI caregiving, with public narratives supporting relational AI [54,97]. |
| Sweden | 7 | Secular, posthumanist leanings support emotional AI, although philosophical critiques remain [142]. |
| UAE | 5 | National AI ambition is high, but traditional religious views challenge affective AI roles [24,151]. |
| United Kingdom | 8 | Human rights discourse and digital pluralism support relational AI in public services [135,145]. |
| United States | 6 | AI companions are culturally accepted, but moral concerns from conservative groups temper deeper emotional integration [23,97]. |

### 4.1.3 Dimension 3: Legal, Ethical & Consent Frameworks

As AI systems increasingly interact with human users in emotionally intimate settings—such as caregiving, companionship, and education—their governance must extend beyond conventional privacy or data norms. This dimension evaluates how national frameworks regulate emotional AI, focusing on affective consent, trust safeguards, and ethical accountability.

Traditional regulatory tools often overlook the risks specific to relational AI: emotional manipulation, dependency, or ambiguous agency in child-AI or elder-AI interactions. Scholars like [152] and [92] argue that this calls for a shift from agential regulation to relational ethics, where symbolic trust and shared responsibility guide system design and deployment.

Furthermore, ethical models must be culturally sensitive, emphasizing human dignity in vulnerable contexts [47,140]. The presence of transparent, dynamic consent mechanisms is especially critical for users such as children, elders, or persons with cognitive impairments.

AFII's third dimension complements the AI Index's policy and governance indicators [77], extending them by asking: Can the state not only regulate AI use, but also safeguard affective safety in familial settings?

Please refer Table 7 for scoring criteria and Table 8 for comparative country scores.

| Table 7: Criteria of Measures for Legal, Ethical & Consent Frameworks | | |
|---|---|---|
| Score Range | AFI Readiness Aspects | Key References |
| 1–2 | - No legal recognition of AI systems<br>- No consent frameworks for emotional AI<br>- Emotional manipulation unpunished | [153]; [15] |
| 3–4 | -Non-binding ethics guidelines<br>- No explicit regulation for emotional safety<br>- Informal consent norms | [24];<br>[154] |
| 5–6 | - Generic AI policies with no focus on emotional domains<br>- Informed consent present but affective dimensions excluded<br>- Fragmented oversight | [152]; [155]; [122] |
| 7–8 | - Emotional AI clauses exist in health or education<br>- Sectoral regulation with dynamic consent trials<br>- Enforcement agencies monitor public use | [36];<br>[46] |
| 9–10 | - Dynamic, legally binding consent frameworks<br>- Emotional AI regulated across all domains<br>- Protocols for trust audits, safety, and explainability. | [46]; [54];[152] |

Countries such as Singapore, Sweden, Germany, and South Korea lead in embedding emotional AI within legal and ethical systems—offering dynamic consent, explainability mandates, and affective safety protocols [118,128]. In contrast, mid-tier nations like Japan, the U.S., China, and UAE have made regulatory advances but still lack comprehensive consent frameworks in caregiving and familial domains [46,77]. Meanwhile, India, Brazil, and South Africa are developing national AI ethics charters, yet enforcement gaps and limited attention to emotional safety restrict their readiness [12,156].

This dimension thus evaluates whether a society recognizes AI not merely as an operational tool, but as a relational presence—one that demands safeguards of trust, dignity, and consent in emotionally embedded contexts. (See Table 8 for detailed scores and justifications.)

| Country | Score * | Justification |
|---|---|---|
| | | **Table 8: Country-Level Scores on Legal, Ethical & Consent Frameworks and with Justification** |
| Brazil | 5 | National ethics guidelines exist, but emotional AI remains unregulated in caregiving or family contexts [12,14]. |
| China | 7 | Active regulation in health and education sectors; emotional consent is top-down with limited individual autonomy [126,127,157]. |
| France | 8 | Benefits from GDPR and upcoming EU AI Act; emotional AI in public systems includes consent and safety provisions [21,128]. |
| Germany | 9 | Legally binding emotional AI regulation aligned with EU law; strong emphasis on explainability and relational ethics [128,129]. |
| India | 6 | AI ethics included in national strategy, but lacks formal affective consent structures. Progress is underway [36,156]. |
| Japan | 8 | Government initiatives support ethical caregiving AI; emotional consent practices are emerging in health tech [46,123]. |
| Singapore | 9 | Comprehensive governance framework includes affective safety, consent-by-design, and explainability standards [24,118]. |
| South Africa | 5 | Ethical boards exist but emotional AI is largely unregulated. National strategy still evolving [131,132]. |
| South Korea | 9 | National policy supports caregiving robots with affective safeguards; consent is legally recognized [54,130]. |
| Sweden | 9 | Enforces GDPR and the EU AI Act, ensuring robust emotional safety and transparency in AI systems [128,129]. |
| UAE | 7 | AI strategy includes ethics, but emotional AI governance is emerging. Guidelines are in development [24,158,159]. |
| United Kingdom | 9 | Emotional AI is integrated into public-sector frameworks, supported by consent-by-design and trust audits [135,160]. |
| United States | 7 | The AI Bill of Rights and NIH initiatives promote ethical AI, though emotional consent standards remain fragmented [111,137]. |

*- Scores are derived from literature review and quantitative data from Stanford HAI (2024) and Tortoise (2024), recalibrated through this study's AFII criteria measures. [77] [133]

### 4.1.4 Dimension 4: Local Adaptability & Inclusivity

This dimension evaluates a country's ability to embed relational and emotional AI systems across its cultural, linguistic, geographic, and socio-economic diversity. It reflects whether AI tools are designed for inclusion, rather than reinforcing structural disparities.

As emotional AI enters roles such as child tutoring, elder companionship, and disability support, it must be adapted to local caregiving values, family norms, and vernacular languages. High-performing countries ensure that AI tools are not only accessible but also culturally resonant and symbolically equitable—serving minority households, indigenous communities, and rural regions through adaptive design and inclusive policy [46,131,161].

Scholars warn of the risks of symbolic exclusion. If AI systems only reflect urban, elite, or majority cultural norms, they risk deepening emotional disenfranchisement [152,162]. Others emphasize the need for co-designed systems that reflect caregiving diversity and do not reinforce gendered or class-based emotional labor [58,125].

Aligned with the Policy and Governance and Public Perception pillars of the AI Index [77], this dimension moves beyond infrastructure to evaluate a society's ability to localize emotional AI meaningfully and equitably. See Table 9 for scoring indicators and Table 10 for country-specific performance.

| Table 9: Criteria of Measures for Local Adaptability & Inclusivity | | |
|---|---|---|
| Score Range | AFI Readiness Aspects | Key References |
| 1–2 | - No localization; urban-elite focused AI only<br>- Ignores regional, linguistic, or cultural variation<br>- No access for rural or disabled populations | [161]; [162]; |
| 3–4 | - Basic translations or regional pilots<br>- AI concentrated in tech hubs; symbolic exclusion persists<br>- Limited public access | [120];[119] |
| 5–6 | - National-level efforts to reach multiple linguistic zones<br>- Inclusive access uneven; disability and tribal regions underserved<br>- Partial efforts in emotional co-design | [58];[52] |
| 7–8 | - AI tools in local languages and schools<br>- Targeted programs for rural, aging, and minority families<br>- Gender-responsive and accessibility-aware designs | [46]; [125] |
| 9–10 | - Policy mandates emotional AI inclusion across regions<br>- Strong co-design culture, multicultural AI personas<br>- Systemic inclusivity in symbolic, ethical, and technical design | [131];[118] |

As shown in Table 10, Singapore, Japan, Sweden, and Germany lead with systemic approaches to AI inclusion—combining emotional design, linguistic adaptability, and regional outreach. Their strategies reflect a deep cultural readiness to accept AI as an emotionally intelligent presence across diverse household structures and social identities.

In contrast, countries like India, China, the UAE, and the United States show strong momentum in infrastructure and policy but still face notable gaps in reaching rural, indigenous, and underserved communities. Despite technological advancement, emotional co-design and symbolic inclusivity remain fragmented.

Meanwhile, South Africa and Brazil face more foundational challenges: while pilot programs exist, national-level commitment to equitable emotional AI access is inconsistent and underdeveloped.

Ultimately, this dimension reveals whether AI is being developed for the many, not just the elite—and whether it can truly resonate across lines of culture, disability, geography, and care ethics [152] [131] [125].

**Table 10: Country-Level Scores on Local Adaptability & Inclusivity with Justification**

| Country | Score* | Justification |
|---------|--------|---------------|
| Brazil | 5 | Urban pilot programs active, but indigenous and rural communities remain excluded from emotional AI [14,120]. |
| China | 7 | Geographic AI access expanding, but top-down design limits cultural inclusivity [126,127,157]. |
| France | 8 | Multilingual integration in education and healthcare; policies support regional diversity [21,128]. |
| Germany | 8 | Strong inclusion policies for migrants, elderly, and disabled; inclusive healthcare AI initiatives [129]. |
| India | 6 | AI policy recognizes regional diversity, but emotional co-design and tribal inclusion are limited [119,163]. |
| Japan | 9 | Aging-focused robots adapted to regional norms; cultural co-design embedded in deployment [46,123]. |
| Singapore | 10 | Comprehensive AI integration in public services, eldercare, and disability support; multilingual, inclusive by design [24,118]. |
| South Africa | 4 | Urban inclusion efforts underway, but rural emotional access and cultural AI design remain weak [131,132]. |
| South Korea | 8 | AI tools serve aging, rural, and disabled groups; caregiving design reflects regional needs [54,130]. |
| Sweden | 9 | Inclusive AI deployment in healthcare and education with multicultural and gender-aware design [128,129]. |
| UAE | 7 | Multilingual and education-focused AI expanding, but outreach to rural/migrant communities remains in progress [24]. |
| United Kingdom | 8 | NHS and public sector AI address disability and gender diversity; regional adaptability increasing [135,160]. |
| United States | 6 | Equity pilots and bilingual AI exist, but emotional inclusion across rural and marginalized groups remains fragmented [77,164]. |

*- Scores are derived from literature review and quantitative data from Stanford HAI (2024) and Tortoise (2024), recalibrated through this study's AFII criteria measures. [77] [133]

### 4.1.5   Dimension 5: AI Talent, Youth Exposure & Emotional Literacy

This dimension examines a society's capacity to equip its youth with both technical fluency and emotional intelligence in relation to AI systems. It evaluates whether educational ecosystems prepare young people not only to build AI, but to ethically co-exist with it— understanding its relational, affective, and social impact.

Key indicators include:

- Integration of AI education across primary to tertiary levels
- Programs for emotional AI co-design, ethics, and caregiving scenarios
- Youth exposure to relational AI in classrooms, labs, or everyday life

Countries that foster early, inclusive AI engagement—through coding curricula, hackathons, carebot initiatives, and empathy-based modules—demonstrate stronger preparation for emotionally literate AI citizenship [165–167]. These programs help students set healthy boundaries with AI companions, navigate affective technologies, and design ethically conscious systems.

As [163,166] argues, youth-facing AI must be grounded in affective scaffolding, avoiding empathy illusions or manipulative feedback loops. Others, like [61] and [168], highlight the value of intergenerational AI education—teaching young users to co-create emotionally safe, socially contextualized AI.

Governments that invest in relational AI education—such as Singapore's AI4KIDS, South Korea's carebot labs, and Germany's ethics modules—are laying the groundwork for AI systems embedded not just in devices, but in values.

AFII Dimension 5 aligns with the Economy and Education pillars of the AI Index [77], tracking AI curriculum, emotional learning, and youth engagement in co-design.

See Table 11 for scoring criteria and Table 12 for national readiness levels.

| Table 11: Criteria of Measures for AI Talent, Youth Exposure & Emotional Literacy | | |
|---|---|---|
| **Score Range** | **AFI Readiness Aspects** | **Key References** |
| **1–2** | - No AI curriculum or exposure<br>- AI seen as elite tool, not relational<br>- No ethical or emotional training | [167];[166];<br>(Ahmad et al., 2023);[170] |
| **3–4** | - Limited exposure in urban schools<br>- Focus on STEM, not empathy or ethics<br>- No emotional literacy integration | [165]; [61]; [171] |
| **5–6** | - AI in general curriculum; coding-focused<br>- Emotional/relational aspects not taught<br>- Informal learning via apps or games | [172]; [119];(Ahmad et al., 2022) |
| **7–8** | - National AI initiatives include emotional literacy<br>- Youth co-design, AI ethics, caregiving bots included<br>- Workshops, labs, and policy support | [129]; [135] [163] |
| **9–10** | - Mandated AI emotional design education<br>- Early exposure to relational AI<br>- Ethical reasoning, empathy, and symbolic literacy taught | [118]; [54]; [46]; [61] |

Countries like Singapore, Japan, South Korea, and China lead in fostering emotionally literate AI education, combining early exposure with robust training in ethical reasoning and relational understanding. Their programs prepare youth not only to build AI systems, but to engage with them as affective and social companions.

France, Germany, Sweden, and the United Kingdom follow closely, showing strong integration of AI ethics and emotional design into national curricula. In contrast, India, the UAE, and the United States are progressing, but face gaps in access and consistency—particularly in emotional or caregiving contexts.

Brazil and South Africa remain in early developmental phases, with promising pilot programs but limited systemic adoption of relational AI education.

This dimension ultimately measures how a society invests in its next generation's capacity to form ethically grounded, emotionally aware relationships with AI—laying the foundation for future AI-family cohabitation that is both technically enabled and socially meaningful [61,165,166]. (See Table 12)

**Table 12: Country-Level Scores on AI Talent, Youth Exposure & Emotional Literacy with Justification**

| Country | Score* | Justification |
|---------|--------|---------------|
| Brazil | 6 | AI curriculum introduced in public urban schools, but emotional AI education remains underdeveloped and rural inclusion is minimal (Ahmad et al., 2022; Smriti, 2024). |
| China | 9 | Emotional AI education integrated in provincial pilots, including ethics and tutoring robots; co-design in classrooms emerging [54,126]. |
| France | 7 | AI integrated in high school curricula; emotional ethics components under development [21,128]. |
| Germany | 8 | National programs emphasize trustworthy AI and integrate ethics workshops into school systems [46,129]. |
| India | 7 | CBSE curriculum includes AI and empathy learning, but wide disparities persist for rural and tribal youth [119,125,163]. |
| Japan | 9 | Carebot use in schools is normalized; cultural acceptance facilitates widespread emotional AI exposure at all education levels [46,123]. |
| Singapore | 10 | AI education begins in primary school; emotional literacy and relational ethics are embedded in national frameworks [24,123]. |
| South Africa | 5 | Urban pilots exist, but lack consistent national AI curriculum and emotional design exposure [131,132]. |
| South Korea | 9 | AI education includes caregiving robotics, youth co-design, and emotional ethics in school curricula [54,130]. |
| Sweden | 8 | Emotional design and coding integrated across education system, supported by national AI strategy [128,129]. |
| UAE | 7 | Schools and universities include AI education; emotional literacy components are emerging but not yet universal [24]. |
| United Kingdom | 8 | Coding and emotional AI integrated in national curriculum; Turing Institute supports empathy-oriented projects [135,160]. |
| United States | 7 | Strong AI education presence, but emotional AI exposure varies significantly across states and institutions [137,172]. |

*- Scores are derived from literature review and quantitative data from Stanford HAI (2024) and Tortoise (2024), recalibrated through this study's AFII criteria measures. [77] [133]

### 4.1.6   Dimension 6: Social Narrative & Symbolic Trust

As AI enters intimate spaces—homes, schools, and care settings—public trust becomes more than a question of technical accuracy or legal protection. It is also emotional and symbolic, shaped by cultural narratives, storytelling, and the way AI is framed in media, education, and policy discourse.

High-scoring countries in this dimension cultivate emotionally resonant, culturally attuned portrayals of AI, helping normalize it as a relational presence. These societies leverage storytelling, arts, and ethical campaigns to bridge the gap between technical feasibility and social acceptance [61,168].

Narrative trust emerges through:

- Media depictions that humanize AI while teaching critical literacy—encouraging thoughtful emotional engagement without naïve overtrust [46,97].
- Public campaigns that emphasize empathy, care, and relational ethics over productivity or control [24].
- Academic and policy narratives that frame AI as part of moral and social life, not just technological infrastructure [162].

Scholars like [163,166] show how narratives about "AI as kin" shape how children and elders relate to AI, influencing attachment, agency, and even trauma processing. Yet caution is essential: as [173] argue, uncritical anthropomorphism may result in emotional overreach, blurring lines between companionship and control.

Symbolic trust is also culturally mediated. As [46] and [174] explain, religious and philosophical values strongly influence whether AI is seen as a legitimate caregiver or ethical agent. Confucian and animist cultures often embrace AI as relational; Western legalistic traditions tend to be more reserved.

Ultimately, this dimension (see Table 13) reflects how societies imagine and narrate AI—whether they prepare citizens to welcome AI as a relational figure responsibly, or treat it merely as a tool. High performers strike a delicate balance: building emotional accessibility without erasing ethical boundaries.

| Table 13: Criteria of Measures for Social Narrative & Symbolic Trust | | |
|---|---|---|
| Score Range | AFI Readiness Aspects | Key References |
| 1–2 | - AI framed as threat or alien<br>- No positive relational AI discourse<br>- Symbolic resistance or fear dominates | [168];[166]; [122] |
| 3–4 | - AI seen as tool for productivity<br>- No emotional or caregiving roles represented<br>- Little to no public education efforts | [137];[15] |
| 5–6 | - Media occasionally portrays AI as tutor, assistant<br>- Some public awareness campaigns<br>- Limited diversity in symbolic representation | [61];<br>[172] |
| 7–8 | - AI framed as emotionally capable but with caution<br>- Media and government highlight relational ethics<br>- Youth and elder acceptance promoted | [46];[135];[163] |
| 9–10 | - AI symbolized as caring, safe, and emotionally intelligent<br>- Integrated into cultural narratives of family and care<br>- Symbolic boundaries and critical thinking also taught | [118];<br>[54];<br>[168];[123] |

As shown in Table 14, countries like Singapore, Japan, Sweden, South Korea, and Germany lead in fostering symbolic trust in AI. Through culturally grounded narratives that emphasize emotional intelligence, caregiving roles, and ethical restraint, these societies cultivate a form of critical trust—welcoming AI into intimate spaces without erasing symbolic boundaries.

France, the United Kingdom, China, and the United States are advancing in symbolic integration, portraying AI as assistants or tutors in public discourse. However, these representations often lack emotional nuance, especially across marginalized or intergenerational groups.

In contrast, India, Brazil, and South Africa are still shaping their narrative ecosystems. While awareness is growing, symbolic trust remains fragmented—framed more through utility than relational legitimacy, and with limited emotional AI literacy.

Ultimately, this dimension captures whether a society is prepared to see AI not just as a tool, but as a symbolic companion—one that deserves emotional responsibility, ethical scrutiny, and cultural care.

| Table 14: Country-Level Scores on Social Narrative & Symbolic Trust with Justification | | |
|---|---|---|
| Country | Score* | Justification |
| Brazil | 5 | AI framed mostly as educational tool; symbolic and emotional diversity in narratives remains low [14,124]. |
| China | 7 | State-driven symbolic trust supports AI in caregiving roles, but emotional framing remains centralized [126,127]. |
| France | 8 | Cultural narratives present AI as ethical companion; symbolic boundaries are actively discussed [21,128]. |
| Germany | 8 | Public discourse emphasizes assistive AI and relational trust with clear ethical framing [129,168]. |
| India | 6 | Youth-facing campaigns exist, but portrayals are often utilitarian and lack emotional nuance [119,166]. |
| Japan | 9 | Longstanding cultural portrayal of AI as caregiver and emotional partner; deeply embedded in media [46,123]. |
| Singapore | 10 | AI is symbolically framed as caring, safe, and emotionally intelligent through national campaigns [24,118]. |
| South Africa | 5 | Public understanding of AI is growing, but narratives remain technocratic or elite-focused [131,132]. |
| South Korea | 9 | Media and policy reinforce symbolic trust in caregiving AI; supported by education and public engagement [54,130]. |
| Sweden | 9 | Inclusive and emotionally nuanced AI narratives promoted through public media and ethics education [128,129]. |
| UAE | 7 | National strategy improves symbolic trust, but emotional portrayals remain concentrated in elite sectors [24]. |
| United Kingdom | 8 | Turing-led initiatives promote relational AI literacy; symbolic balance emphasized in curricula [135,160]. |
| United States | 7 | AI is culturally visible as assistant and threat; symbolic trust varies across communities and platforms [137,168]. |

*- Scores are derived from literature review, recalibrated through this study's AFII criteria measures.

### 4.1.7 Dimension 7: Historical Adoption & Industry Leadership

This dimension evaluates a country's long-term engagement with AI, focusing on how deeply AI has been adopted across institutional, industrial, and family-relevant sectors such as healthcare, education, eldercare, and emotional wellbeing. It considers not only technical leadership and research capacity, but also the extent to which AI innovation has supported relational, affective, and caregiving roles.

High-performing nations in this category are characterized by decades of R&D investment, sustained public-private collaboration, and ethical leadership in AI design and deployment. Their histories in AI adoption allow for smoother integration of emotionally intelligent systems within everyday caregiving and educational contexts [24,152].

As [128]and others note, such leadership often includes pioneering work on AI governance, relational safety protocols, and emotion-aware design standards. However, as [166] and [36] emphasize, early adoption alone does not guarantee readiness. Emotional and ethical contextualization is essential to transition from technological maturity to relational integration.

This dimension also captures industry foresight—the ability to lead on affective computing, emotion-AI co-regulation, and public engagement through global platforms like AI Expo Korea (2025) or initiatives under UNESCO and GPAI. National strategies that explicitly include emotional intelligence, caregiving applications, and family-facing use cases are strong indicators of readiness.

Aligned with the R&D, Economy, and Education pillars in the AI Index [77], this dimension rewards not just innovation, but its ethical depth and emotional relevance over time. See Table 15 for scoring indicators and Table 16 for country-level evaluations.

**Table 15: Criteria of Measures for Historical Adoption & Industry Leadership**

| Score Range | AFI Readiness Aspects | Key References |
|---|---|---|
| 1–2 | - No AI industry or research capacity<br>- No history of AI in health or education<br>- Minimal public-private collaboration | [121]; [131] |
| 3–4 | - AI startup scene emerging<br>- Focus limited to finance or automation<br>- No relational AI investment | [12];[175] |
| 5–6 | - AI adopted in public and commercial sectors<br>- Moderate R&D ecosystem<br>- Family-facing use cases minimal | [124]; [119]; [36] |
| 7–8 | - Significant AI adoption in health and education<br>- AI ethics research growing<br>- Youth and elder-focused AI innovations present | [135]; [66]; [163] |
| 9–10 | - Global AI leadership<br>- Longstanding R&D investment<br>- Active leadership in caregiving, emotional and symbolic AI innovation | [128];[24];[118]; [54] |

As shown in Table 16, Singapore, the U.S., Japan, Germany, and South Korea lead with decades of sustained AI investment, R&D leadership, and consistent innovation across relational domains—particularly caregiving, mental health, and education.

France, Sweden, the UK, China, and India demonstrate strong national AI trajectories, but differ in how fully their ecosystems address the emotional depth and public accessibility required for AI-family integration.

In contrast, Brazil, South Africa, and the UAE are gaining momentum through national strategies and pilot initiatives, yet still lack the historical grounding and sectoral integration needed to embed AI meaningfully in caregiving or emotional contexts.

This dimension ultimately assesses whether a nation's AI journey is not only technically advanced, but also ethically driven and relationally capable—key factors for long-term AI-family readiness.

| Country | Score* | Justification |
|---|---|---|
| **Table 16: Country-Level Scores on Historical Adoption & Industry Leadership and with Justification** | | |
| Brazil | 4 | Recent activity in AI startups and edtech, but limited historical engagement or emotional AI investment [12,124] |
| China | 9 | Leader in robotics, AI infrastructure, and smart cities; strong adoption in education and health AI [126,176,177] |
| France | 8 | The Villani Report advanced national AI strategy; ethics and relational AI supported through R&D [21,128] |
| Germany | 9 | EU AI policy leader with deep integration in eldercare and education; consistent investment in ethics and affective computing [128,129] |
| India | 7 | Rapidly growing AI ecosystem; early AI education and edtech adoption, though emotional AI remains nascent [119,166] |
| Japan | 9 | Pioneer in caregiving robotics and aging-focused AI; relational and emotional AI part of long-term national strategy [46,123] |
| Singapore | 10 | World leader in AI policy integration; Smart Nation framework emphasizes emotional design, ethics, and cross-sector AI use [24,118] |
| South Africa | 4 | AI innovation ecosystem still developing; limited engagement with relational or emotional domains [131,132] |
| South Korea | 9 | Government-led investment in robotics and relational AI; strong public-private caregiving AI collaborations [54,130] |
| Sweden | 8 | Long-standing AI research institutions; contributes to ethical frameworks and relational AI innovation in the EU [128,129] |
| UAE | 7 | Strong national AI strategy and growing interest in health AI; emotional integration still emerging [24,175] |
| United Kingdom | 9 | Home to Turing Institute and AI ethics research hubs; NHS piloting AI in relational and caregiving contexts [135,160] |
| United States | 10 | Early AI pioneer with unmatched investment and innovation across education, health, and relational AI domains [66,77,137] |

*- Scores are derived from literature review and quantitative data from Stanford HAI (2024) and Tortoise (2024), recalibrated through this study's AFII criteria measures. [77] [133]

### 4.1.8   Dimension 8: Family Structure & Emotional Labor Equity

This dimension (Table 17) evaluates how a country's family norms, caregiving patterns, and gendered distribution of emotional labor influence its capacity to integrate relational AI into caregiving and intimate domestic roles. Family structures—whether nuclear, multigenerational, extended, or care-fragmented—shape how relational AI will be culturally received and practically deployed as a companion, caregiver, or emotional surrogate.

In societies where caregiving is feminized, undercompensated, or socially invisible, there is a risk that AI will reinforce rather than alleviate these inequities. [43] critiques the way AI is embedded within historically gendered care economies, warning against technological "patchwork" that sustains emotional labor imbalances. [178] and [47] similarly argue that unless AI systems are ethically and structurally designed to challenge these disparities, they may serve as digital proxies for unpaid or undervalued care work.

Countries scoring higher in this dimension demonstrate:

- Policy frameworks that explicitly address care labor equity
- Gender-inclusive caregiving models
- Co-design of AI tools with diverse family configurations and care burdens in mind

[61] emphasize the importance of early educational exposure to AI ethics and emotional labor for children, shaping future users and designers of empathetic AI systems. [166] calls for child-safe, relationally intelligent AI design that reflects awareness of gender and care work inequalities, particularly in emotional surrogate roles. [162] highlight how culturally specific caregiving norms must be embedded into AI design, ensuring resonance with plural domestic values and care traditions.

Furthermore, [120] and [125] point out that relational AI, if uncritically designed, may encode dominant caregiving scripts that marginalize alternative family structures, such as queer kinship, single parenting, or intergenerational cohabitation. Equity in emotional labor thus requires inclusive design—not only at the technological level, but within the broader socio-political recognition of diverse caregiving actors and expectations.

This dimension (refer Table 10), therefore, does not only assess a country's AI infrastructure or ethics—it gauges whether the emotional ecosystems of care in that society are capable of responsibly hosting AI as a relational ally, rather than a convenient but exploitative stand-in.

| Table 17: Criteria of Measures for Family Structure & Emotional Labor Equity | | |
|---|---|---|
| Score Range | AFI Readiness Aspects | Key References |
| 1–2 | - High gendered division of care<br>- No public care infrastructure<br>- AI seen as housework replacement | [121]; [120] |
| 3–4 | - Informal caregiving common<br>- No policies on care equity<br>- Relational AI framed as "helper" for women | [125]; [122] |
| 5–6 | - Public discourse on emotional labor growing<br>- Some programs for shared caregiving<br>- Relational AI introduced but not regulated | [166]; [162] |
| 7–8 | - National strategies include gender/care equity<br>- Emotional labor audits or training exist<br>- Family-inclusive AI design practices encouraged | [162]; [61] |
| 9–10 | - Institutional caregiving policies embedded<br>- Gender- and care-sensitive AI co-design mandated<br>- AI used to reduce domestic emotional load equitably | [118]; [46] |

Singapore, Sweden, Germany, France, and South Korea show high readiness by aligning AI deployment with care equity principles, including support for shared emotional labor and family-inclusive design.

Japan, China, UK, and the U.S. exhibit strong relational AI capacity but must more directly address gendered care structures and emotional labor distribution in domestic settings.

India, Brazil, UAE, and South Africa show growing access to AI tools but lack institutional frameworks to ensure AI enhances—rather than exploits—existing family caregiving dynamics.

This dimension reveals whether AI is introduced into homes as a relational partner in care—or simply another layer of invisible labor (refer Table 18).

| Table 18: Country-Level Scores on Family Structure & Emotional Labor Equity with Justification | | |
|---|---|---|
| **Country** | **Score*** | **Justification** |
| Brazil | 5 | Family caregiving burden remains on women; relational AI is emerging but lacks emotional labor framing [14,125] |
| China | 7 | AI integrated into eldercare; traditional filial roles persist; limited policy on gendered emotional labor [126,176] |
| France | 8 | Gender equality in caregiving supported; emotional labor discourse reflected in AI caregiving pilots [21,128] |
| Germany | 8 | Strong parental leave policies and AI integration into care work foster shared emotional labor [128,129] |
| India | 6 | Extended family caregiving common; gendered caregiving persists; AI used in edtech and eldercare) [119,120] |
| Japan | 8 | Advanced carebot tech used in aging society; caregiving seen as communal, though gender roles remain traditional [46,123] |
| Singapore | 9 | Government supports gender-neutral caregiving roles; AI caregiving tools framed as shared labor enhancers [24,118] |
| South Africa | 5 | Care labor is largely informal and female-led; relational AI integration remains underdeveloped [131,132] |
| South Korea | 8 | Carebot policy supports equitable eldercare; emotional labor increasingly recognized in policy [54,130] |
| Sweden | 9 | Global leader in gender-equal parenting and AI integration into public care systems [128,129] |
| UAE | 6 | Domestic care outsourced to migrant labor; AI initiatives emerging, but emotional labor remains unaddressed [24,175] |

| United Kingdom | 8 | Public policy promotes shared care; AI in social services tested for emotional burden reduction [135,160] |
| United States | 7 | AI caregiving innovations exist; gender equity in domestic emotional labor remains inconsistent [122,137] |

*- Scores are derived from literature review, recalibrated through this study's AFII criteria measures.

### 4.1.9 Dimension 9: Economic Accessibility & Equity

This dimension assesses how a society's care norms, family structures, and gendered emotional labor patterns shape its readiness to integrate AI into relational domains such as caregiving, companionship, and domestic support. It evaluates whether AI deployment serves to correct or compound caregiving inequalities.

In settings where emotional labor is feminized, informal, or undervalued, relational AI may risk acting as a technological surrogate, masking deeper systemic imbalances. Scholars like [43] and [178] caution that unless guided by inclusive design and policy, AI may sustain care inequity rather than disrupt it.

Countries scoring highly in this dimension are those with:

- Policy frameworks that address care equity and labor distribution
- AI systems co-designed for diverse family types and caregiving contexts
- Ethical commitments to ensure AI supports, rather than replaces, shared emotional work

As [61] and [166] stress, early exposure to AI ethics and emotional labor is critical for future generations. Others like [120] and [125] argue for AI training models that reflect non-traditional family formations, including queer, single-parent, and intergenerational caregiving dynamics.

This dimension ultimately examines whether a society's emotional care ecosystem is robust enough to host AI as a relational ally, rather than layering automation onto already invisible labor. It links to Economy and Education and Policy and Governance pillars in the AI Index [77], through its focus on access equity, welfare integration, and ethical AI design. See Table 17 for scoring indicators and Table 18 for cross-national performance.

| Table 19: Criteria of Measures for Economic Accessibility & Equity | | |
|---|---|---|
| Score Range | AFI Readiness Aspects | Key References |
| 1–2 | - AI restricted to elite markets<br>- No public access or subsidies<br>- Emotional AI considered luxury tech | [121]; [152] |
| 3–4 | - Urban AI availability only<br>- Relational tools unaffordable for most<br>- No AI equity policy exists | [122]; [120] |
| 5–6 | - Pilot AI projects in schools/clinics<br>- Emotional AI in limited welfare domains<br>- Some subsidies for assistive AI | [166]; [119];[125] |
| 7–8 | - AI embedded in public services<br>- Emotional AI pilots in public schools, hospitals<br>- Income-sensitive AI policies emerging | [24]; [46] |

| | | | |
|---|---|---|---|
| **9–10** | - Emotional AI universally accessible<br>- National funding, open-source emotional AI<br>- Inclusive design + low-cost hardware deployed | [118]; [54]; [131] | |

As outlined in Table 20, Singapore, Sweden, South Korea, Germany, and France lead in ensuring universal or near-universal access to relational AI through public infrastructure. Their policies reflect a high degree of emotional equity planning, embedding AI into caregiving, education, and social welfare systems.

China, Japan, the United Kingdom, and the United States also show strong public-sector AI integration, but continue to face disparities in affordability, regional distribution, and access to emotionally supportive applications, particularly in rural or underserved areas.

Meanwhile, India, Brazil, South Africa, and the UAE are in the early phases of expanding access to relational AI. Despite promising initiatives, structural and economic barriers continue to constrain the equitable deployment of AI as a care-enabling technology.

This dimension ultimately reflects whether AI is evolving as a public good for emotional support—or remains a symbol of elite digital intimacy, accessible only to those with the right location, literacy, or income.

| Table 20: Country-Level Scores on Economic Accessibility & Equity with Justification | | |
|---|---|---|
| **Country** | **Score*** | **Justification** |
| Brazil | 5 | AI tools used in public schools, but access remains urban and income-bound; no national equity strategy [124,125] |
| China | 7 | AI integrated in public education and health, but availability varies across provinces; affordability improving [126,127] |
| France | 8 | Government supports equitable AI distribution in schools and welfare sectors; cost barriers minimal [21,128] |
| Germany | 9 | AI used in inclusive education and eldercare; cost is absorbed by social systems [128,129] |
| India | 6 | Low-cost AI tools introduced in schools; regional disparities in access persist; public sector integration growing [119,166] |
| Japan | 8 | AI caregivers and tutors used in public hospitals and schools; affordability supported through local innovation [46,123] |
| Singapore | 10 | Smart Nation initiative ensures AI access across public housing, education, and welfare; emotional AI available to all [24,118] |
| South Africa | 5 | Urban AI use rising; rural and low-income households under-resourced; pilot access efforts under way [131,132] |
| South Korea | 9 | Emotional AI integrated in eldercare and schools; public subsidies and inclusion mandates active [54,130] |
| Sweden | 9 | Emotional AI available through state welfare and education systems; economic barriers minimal [128,129] |
| UAE | 6 | AI development advanced, but access limited to institutional or private care; migrant and rural equity lacking [24,175] |

| United Kingdom | 8 | NHS and school-based AI pilots ensure moderate accessibility; affordability concerns under review [160,175] |
| United States | 7 | Strong AI innovation, but emotional tools remain income-stratified; federal equity initiatives developing [122,137] |

*- Scores are derived from literature review and quantitative data from Stanford HAI (2024) and Tortoise (2024), recalibrated through this study's AFII criteria measures. [77] [133]

### 4.1.10 Dimension 10: Emotional Authority & Safety Design

This dimension evaluates how well a country anticipates and regulates the emotional influence of relational AI—particularly as such systems assume caregiving, tutoring, or companionship roles. As emotionally expressive agents enter homes, classrooms, and care facilities, they acquire symbolic authority, shaping user mood, behavior, and attachment—especially among children, elders, and emotionally vulnerable populations.

High-performing countries in this domain:

- Regulate emotional simulation and anthropomorphism
- Require affective transparency and user interpretability in design
- Implement co-regulatory frameworks and emotional safety audits for domestic and caregiving AI

Researchers such as [168] and [163,166] warn of emotional overreach—where human-like AI may foster dependency, trust displacement, or false intimacy. Others, including [46] and [61], advocate for relational safety standards, ensuring that emotional AI does not simulate empathy deceptively or manipulate attachment without consent.

The concept of "explainable emotion" [152] reinforces this, calling for systems that reveal their emotional logic and interaction history, reducing ambiguity for users and caregivers. Design that acknowledges symbolic boundaries, especially in sensitive settings like healthcare or education, is central to safe emotional AI deployment.

This dimension ultimately gauges whether emotional AI is understood and governed not only as a technical tool, but as a symbolically potent relational agent—demanding transparency, moral accountability, and cultural restraint. See Table 21 for scoring criteria and Table 22 for comparative country performance.

| Table 21: Criteria of Measures Emotional Authority & Safety Design | | |
|---|---|---|
| **Score Range** | **AFI Readiness Aspects** | **Key References** |
| **1–2** | - No regulation of AI emotion<br>- AI simulates care without limits<br>- No symbolic boundaries on AI intimacy | [168]; [15] |
| **3–4** | - Emotion-AI in public but unregulated<br>- AI mimics attachment without consent<br>- No "empathy disclosure" norms | [122]; [66] |
| **5–6** | - AI design includes soft guidelines<br>- Basic explainability protocols present<br>- Relational misuse not addressed | [152]; [166] |
| **7–8** | - Emotional co-design and audits introduced<br>- Emotional AI evaluated in schools, care sectors<br>- Training for symbolic boundaries exists | [61]; [36]; [46] |
| **9–10** | - National guidelines on emotional safety<br>- Mandatory disclosure for empathic AI<br>- Affective ethics integrated into law and design | [118]; [54]; [160] |

As reflected in Table 22, Singapore, Japan, South Korea, Sweden, Germany, and the United Kingdom lead in regulating emotional AI design, embedding safeguards like transparency, affective explainability, and symbolic boundaries into systems that simulate care, attachment, or empathy.

France, China, and the United States are progressing in relational AI safety, but the absence of mandatory emotional disclosure protocols across all sectors limits comprehensive protection.

By contrast, India, Brazil, the UAE, and South Africa are in early stages of deploying relational AI, with limited symbolic oversight—raising risks of emotional overreach, particularly in caregiving, education, and domestic applications.

This dimension ultimately assesses whether AI is treated not merely as a tool of interaction, but as a relational authority—one that requires ethical governance, emotional transparency, and protective regulation to safeguard against invisible forms of emotional dependency.

**Table 22: Country-Level Scores on Emotional Authority & Safety Design with Justification**

| Country | Score* | Justification |
|---|---|---|
| Brazil | 5 | Emotional AI present in education and services, but lacks clear authority/safety design protocols [14,124] |
| China | 7 | Emotionally expressive AI widely deployed; symbolic disclosure not always practiced; regulations in progress [126] |
| France | 8 | EU AI Act includes emotional safety components; emotional co-regulation models tested in schools [21,128] |
| Germany | 9 | Affective explainability and symbolic literacy part of AI design standards; integrated in public sectors [46,129] |
| India | 6 | Emotional AI present in care and education; safety protocols not formalized; symbolic ethics under development [119,166] |
| Japan | 9 | Long history of relational robotics; safety design emphasized in caregiving bots; emotional boundaries culturally codified [46,123] |
| Singapore | 10 | Affective transparency and co-regulation embedded in national AI design; AI cannot mimic care without disclosure [24,118] |
| South Africa | 5 | Emotional AI pilots exist; lack of national framework for emotional safety or symbolic authority [131,132] |
| South Korea | 9 | Emotional AI used in caregiving with emotional audits, consent, and transparency laws in place [54,130] |
| Sweden | 9 | Strong alignment with EU standards on affective explainability and emotional design ethics [128,134] |
| UAE | 6 | Emotion-AI growing in edtech and care; regulation focuses more on function than emotional transparency [24,175] |

| United Kingdom | 9 | NHS pilots use emotional disclosure protocols; symbolic trust managed in schools and therapy AI [135,160] |
| United States | 8 | AI safety initiatives acknowledge emotional influence; federal protocols remain non-binding [66,137] |

*- Scores are derived from literature review and quantitative data from Stanford HAI (2024) and Tortoise (2024), recalibrated through this study's AFII criteria measures. [77] [133]

### 4.1.11 Real-time AFI Composite Score

The Table 23 below presents composite scores of 13 countries across ten AFI dimensions. These dimensions assess each nation's emotional, ethical, cultural, and infrastructural readiness for AI integration within family and caregiving contexts. Scores are calculated using an equal weight approach, reflecting the relative importance of each dimension in achieving responsible, human-centered AI adoption.

AFII Composite Score = $\frac{1}{10} \sum_{n=1}^{10}$ Dimesnsion(n)

| Table 23: Country-wise Score on AFI Dimensions | | | | | | | | | | | |
|---|---|---|---|---|---|---|---|---|---|---|---|
| | **AFII Dimensions** | | | | | | | | | | **Composite Score (CS)*** |
| **Country** | **1** | **2** | **3** | **4** | **5** | **6** | **7** | **8** | **9** | **10** | |
| | **TIAP** | **CPR** | **LECF** | **LAI** | **ATYEEL** | **SNST** | **HAIL** | **FSELE** | **EAE** | **EASD** | |
| **Brazil** | 5 | 7 | 5 | 5 | 6 | 5 | 4 | 5 | 5 | 5 | 5.2 |
| **China** | 8 | 8 | 7 | 7 | 9 | 7 | 9 | 7 | 7 | 7 | 7.6 |
| **France** | 8 | 6 | 8 | 8 | 7 | 8 | 8 | 8 | 8 | 8 | 7.7 |
| **Germany** | 9 | 6 | 9 | 8 | 8 | 8 | 9 | 8 | 9 | 9 | 8.3 |
| **India** | 6 | 5 | 6 | 6 | 7 | 6 | 6 | 6 | 6 | 6 | 6 |
| **Japan** | 9 | 9 | 8 | 9 | 9 | 9 | 9 | 8 | 8 | 9 | 8.7 |
| **Singapore** | 10 | 8 | 9 | 10 | 10 | 10 | 10 | 9 | 10 | 10 | 9.6 |
| **South Africa** | 4 | 5 | 5 | 4 | 5 | 5 | 5 | 5 | 5 | 5 | 4.8 |
| **South Korea** | 9 | 9 | 9 | 8 | 9 | 9 | 9 | 8 | 9 | 9 | 8.8 |
| **Sweden** | 9 | 7 | 9 | 9 | 8 | 9 | 8 | 9 | 9 | 9 | 8.6 |
| **UAE** | 8 | 5 | 7 | 7 | 7 | 7 | 7 | 6 | 6 | 6 | 6.6 |
| **United Kingdom** | 9 | 8 | 9 | 8 | 8 | 8 | 9 | 8 | 8 | 9 | 8.4 |
| **United States** | 9 | 6 | 7 | 6 | 7 | 7 | 10 | 7 | 7 | 8 | 7.4 |

*CS on equal weight

Table 23 presents the dimension-wise AFII scores and composite scores (CS) for 13 countries, reflecting their relative readiness for emotionally intelligent AI integration in caregiving and family domains. Figure 2 visualizes these scores through a radar chart, highlighting notable divergences—such as Singapore's consistently high performance across all dimensions, contrasted with South Africa and Brazil's lower relational preparedness despite emerging infrastructure. The visual comparison underscores the importance of evaluating AI readiness through caregiving-centric and culturally grounded indicators, as proposed in the AFII framework.

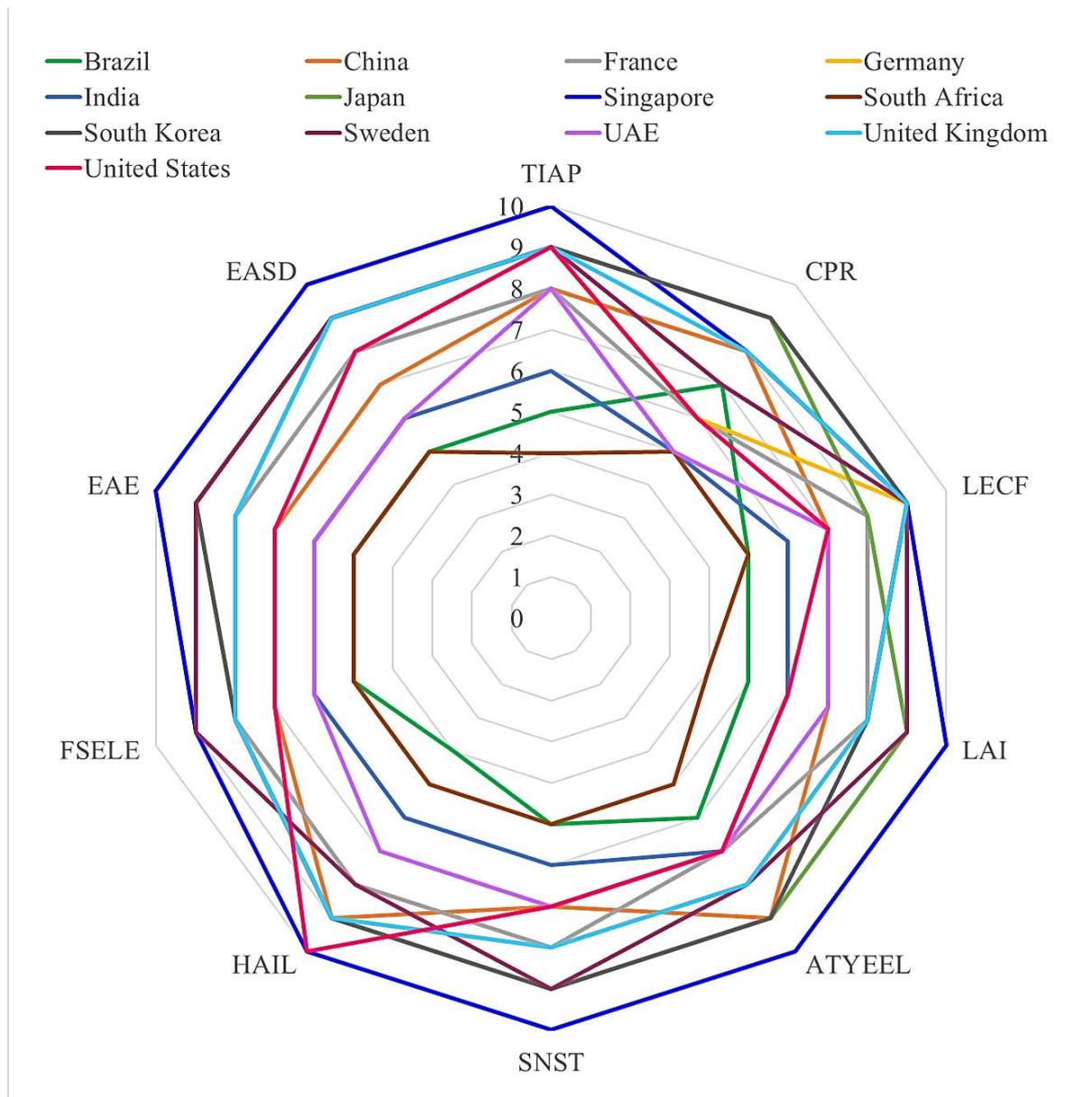

**Figure 2: Comparative Radar Chart of AFII Scores Across 10 Dimensions for 13 Countries**

## 4.2 National Policy Penetration on AFI

As artificial intelligence (AI) systems increasingly engage with the most intimate aspects of human life—such as caregiving, emotional support, and kinship—there is a pressing need to move beyond traditional regulatory priorities. While global AI governance has made significant progress in areas like safety, accountability, and technical reliability, the relational and emotional dimensions of AI–Family Integration (AFI) remain largely underexplored.

In this context, understanding a country's AI Policy Penetration becomes vital. It reflects how deeply AI strategies are embedded across sectors, and whether they consider both technological robustness and ethical sensitivity. The Composite Score (CS) for each country is derived from an equal-weight evaluation of 10 AFI aspects, offering a balanced measure of AI readiness and responsiveness. (Please refer Table 24)

Comparative analysis of these scores reveals global disparities in AI engagement and helps policymakers, educators, and institutions align with emerging standards—highlighting where deeper ethical and human-centered integration is most needed.

| Table 24: Country-wise Composite Score on National Policy Penetration on AI | | | | | | | | | | |
|---|---|---|---|---|---|---|---|---|---|---|
| **Country** | **AFI Aspects / Principles** | | | | | | | | | |
| | **RAI** | **EAI** | **AIC** | **RE** | **AIF** | **ST** | **AK** | **DP** | **AI** | **CAA** | **CS*** |
| Brazil | 5 | 6 | 5 | 4 | 5 | 6 | 4 | 6 | 4 | 6 | 5.1 |
| China | 4 | 3 | 6 | 4 | 4 | 3 | 3 | 5 | 5 | 4 | 4.1 |
| France (EU) | 9 | 10 | 9 | 10 | 9 | 9 | 9 | 10 | 10 | 10 | 9.5 |
| Germany (EU) | 9 | 9 | 9 | 10 | 9 | 10 | 9 | 10 | 10 | 10 | 9.5 |
| India | 6 | 6 | 6 | 6 | 6 | 6 | 6 | 6 | 6 | 6 | 6 |
| Japan | 9 | 10 | 10 | 9 | 10 | 9 | 9 | 10 | 10 | 10 | 9.6 |
| Singapore | 10 | 10 | 10 | 10 | 10 | 9 | 8 | 10 | 10 | 10 | 9.7 |
| South Africa | 3 | 3 | 4 | 4 | 3 | 5 | 3 | 4 | 3 | 4 | 3.6 |
| South Korea | 7 | 9 | 9 | 9 | 8 | 8 | 8 | 9 | 9 | 9 | 8.5 |
| Sweden (EU) | 8 | 9 | 8 | 9 | 9 | 8 | 8 | 9 | 9 | 9 | 8.6 |
| UAE | 6 | 7 | 7 | 7 | 7 | 6 | 6 | 8 | 7 | 7 | 6.8 |
| United Kingdom | 8 | 9 | 9 | 9 | 8 | 9 | 8 | 9 | 9 | 9 | 8.7 |
| United States | 7 | 8 | 8 | 8 | 8 | 8 | 7 | 9 | 8 | 8 | 7.9 |

Note: *CS- Composite Score calculated on equal weightage

**Legend:**
RAI = Relational AI, EAI = Emotional AI, AIC = AI Caregiving, RE = Robot Ethics, AIF = AI Family, ST = Symbolic Trust, AK = AI Kinship, DP = Digital Personhood, AI = AI Intimacy, CAA = Cultural AI Acceptance

**References for Composite Score (CS)**
UNESCO (2021):(UNESCO, 2021a); OECD (2019); [112] EU (France, Germany, Sweden):[114,182]; GPAI (2020): [183]; IEEE (2020): [184]; Brazil: [185]; China:[176]; France: [186]; Germany: [187]; India: [188]. [189] [190]; Japan: [191,192]; Singapore: [193,194]; South Africa: [195]; South Korea: [196,197]; Sweden: [198]; UAE: [199,200]; United Kingdom: [201]; United States (incl. NIST): [202].

Table 24 presents a comparative overview of countries' performance on national policy penetration related to AI, based on the AFI (Artificial Family Intelligence) framework. It evaluates ten core aspects—ranging from ethical AI to cultural acceptance—culminating in a composite score that reflects each nation's policy depth and comprehensiveness. Figure 3, the Geographical Heat Map, visually depicts the global distribution of these composite scores. Darker shades indicate higher policy penetration, offering a quick spatial interpretation of how various countries are progressing in implementing holistic AI policies.

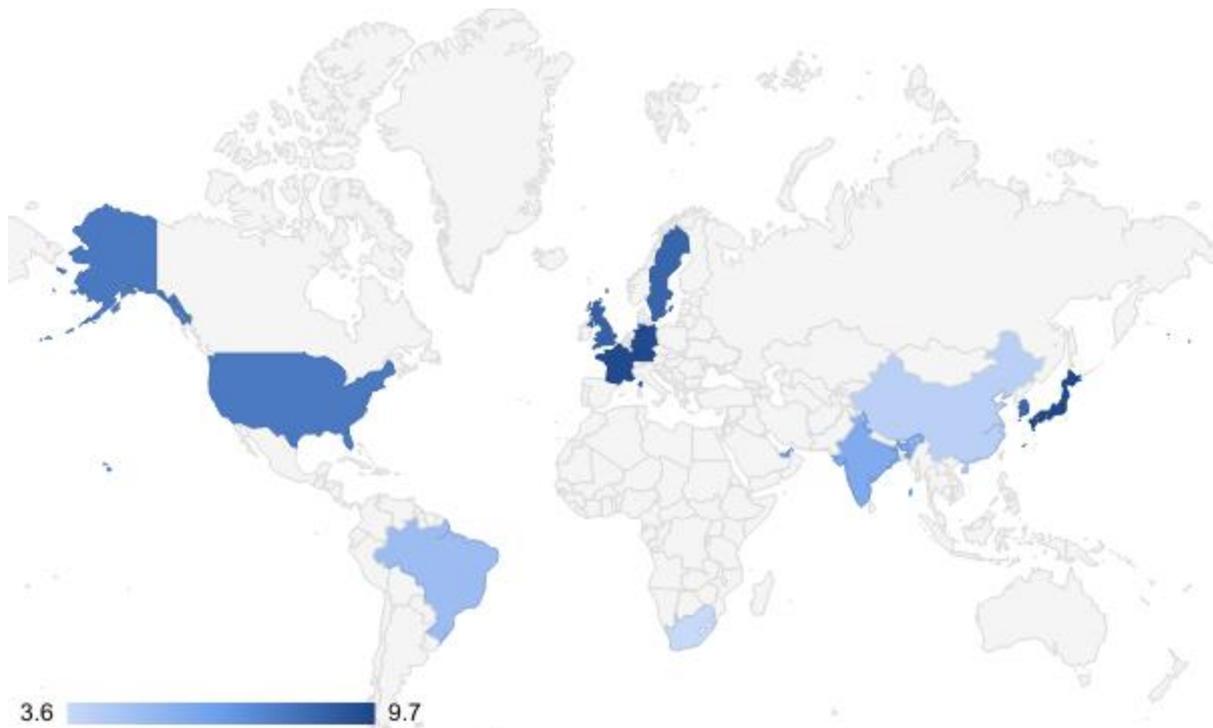

**Figure 3: Geographical Heat Map on AFI Policy Penetration**

## 4.3    Stanford AI Index (2024) – AI Power

The Stanford AI Index [77] is a leading global benchmark published by Stanford University to assess a country's AI development and readiness. It evaluates nations across six key pillars: research & development, talent, infrastructure, economy, governance, and commercialization. These pillars together offer a holistic view of a country's AI ecosystem—from foundational capabilities to real-world impact.

According to the 2024 rankings, the United States holds the top position, followed by China, the United Kingdom, and India, reflecting their leadership in research, innovation, and policy. Other countries like the UAE, France, and South Korea also feature prominently, showing a balanced focus on AI growth. Meanwhile, countries such as Brazil, South Africa, and Sweden are not listed in the top 10, suggesting opportunities to strengthen their AI strategy and global presence (refer Table 25).

| Table 25: Stanford AI Index 2024 | |
|---|---|
| **Country** | **Stanford AI Index 2024\*** |
| United States | 1 |
| China | 2 |
| United Kingdom | 3 |
| India | 4 |
| UAE | 5 |
| France (EU) | 6 |
| South Korea | 7 |
| Germany (EU) | 8 |
| Japan | 9 |
| Singapore | 10 |
| Brazil | Not listed in in Top 10 |
| South Africa | Not listed in in Top 10 |
| Sweden (EU) | Not listed in in Top 10 |

*Based on 6 pillars of AI Index (Stanford HAI, 2024)

# 5.    Insights and Interpretation

## 5.1    Country-wise Real-time Scores and Ranks on AFII Dimensions

The AI-Family Integration Index (AFII) provides a transformative lens for evaluating nations not only on their technological capabilities but also on their preparedness to welcome AI into emotionally sensitive, ethically complex, and culturally rooted caregiving spaces. The AFII's Composite Scores (CS)—ranging from 4.8 to 9.6—reveal compelling contrasts between innovation-driven AI adoption and human-centered readiness.

The graphical visualization and Table 13 highlight both leaders and laggards in this space. While Singapore (CS 9.6) leads globally, nations such as South Korea (8.8) and Japan (8.7) closely follow. The top five countries (Singapore, South Korea, Japan, Sweden, and the UK) all score above 8.4, reflecting high alignment between AI policy, cultural values, emotional literacy, and ethical infrastructure.

In contrast, middle performers like the United States (7.4) and China (7.6), though technologically dominant, exhibit critical shortfalls in symbolic trust, emotional safety, and caregiving readiness. At the lower end, India (6.0), Brazil (5.2), and South Africa (4.8) demonstrate emerging promise but require substantial investment in emotional-AI literacy, inclusive infrastructure, and ethical governance to bridge the gap.

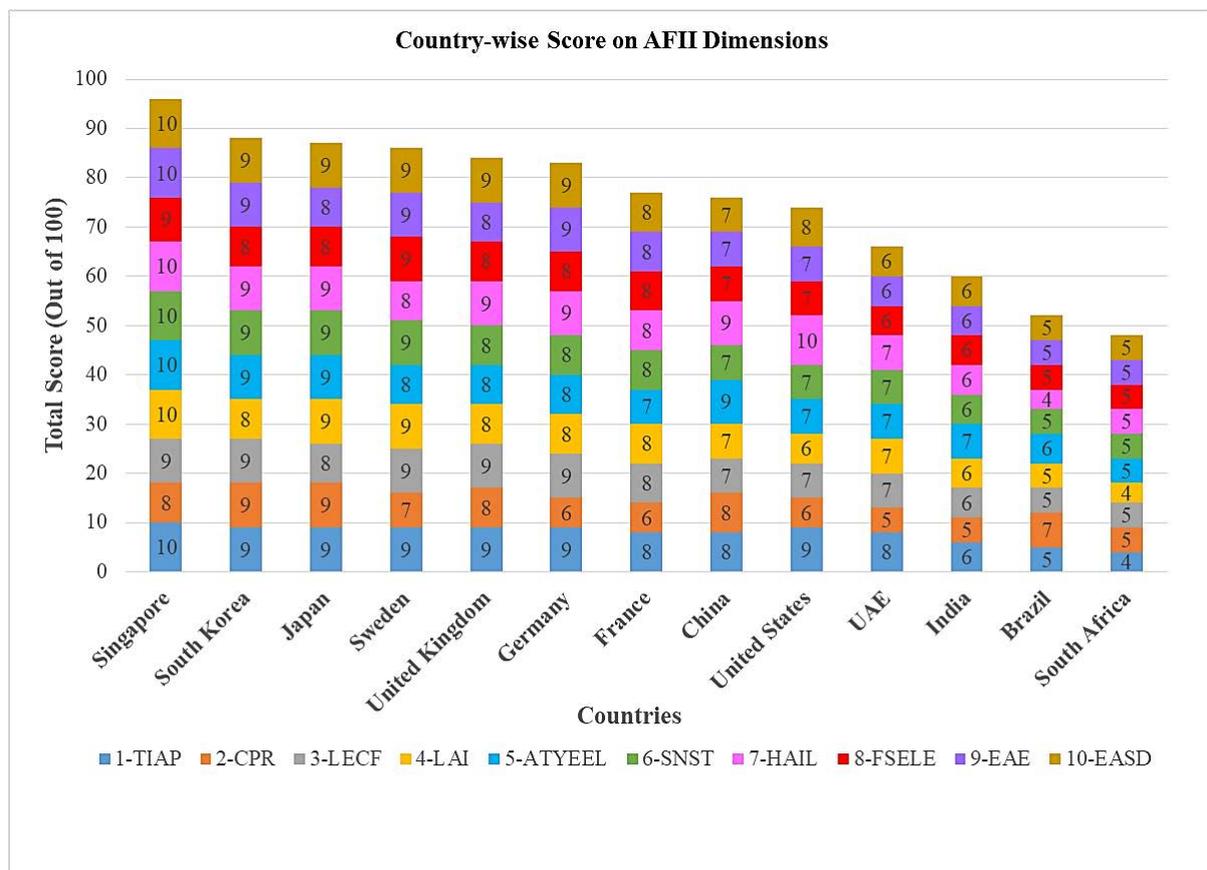

**Figure 4: Country-wise Real-time Scores on AFII Dimensions**

Figure 4 illustrates the country-wise distribution of scores across ten AFII (Artificial Family Intelligence Index) dimensions. Each colored segment represents a specific policy or ethical domain, providing a detailed, dimension-level understanding of national AI readiness and alignment with human-centric values. Table 26 ranks countries based on their composite AFII scores derived from the same ten dimensions. The ranking reflects the overall policy maturity and ethical integration in each country's approach to AI governance.

| Table 26: Country's AFII Rank | | |
|---|---|---|
| **Country** | **Composite Score (CS)\*** | **AFII Rank** |
| Singapore | 9.6 | 1 |
| South Korea | 8.8 | 2 |
| Japan | 8.7 | 3 |
| Sweden | 8.6 | 4 |
| United Kingdom | 8.4 | 5 |
| Germany | 8.3 | 6 |
| France | 7.7 | 7 |
| China | 7.6 | 8 |
| United States | 7.4 | 9 |
| UAE | 6.6 | 10 |
| India | 6.0 | 11 |
| Brazil | 5.2 | 12 |
| South Africa | 4.8 | 13 |

\*- A Composite score obtained by totalling of 10 Dimensions on equal weightage

### 5.1.1 Global Leaders and Their Strategic Strengths

*Singapore: The Relational AI Vanguard*

Singapore (CS 9.6) stands out as the global gold standard in AI-family readiness. Its composite score of 9.6 reflects exceptional performance across all AFII dimensions—particularly in emotional literacy, inclusive design, policy coherence, and caregiving innovation. The nation's integration of AI in early education and eldercare, as well as its focus on emotionally intelligent social robots, aligns with research calling for empathic, human-compatible machines (Abdollahi, 2023; Belpaeme et al., 2018). As AI increasingly enters homes and schools, Singapore's model of care-first AI design sets a powerful precedent.

*East Asia's Empathetic Edge: South Korea and Japan*

South Korea (CS 8.8) and Japan (CS 8.7) are global leaders in affective robotics and relational AI. Both nations exhibit strong performance in legal safeguards (LECF), emotional AI integration (HAIL), and symbolic trust (SNST). Japan's long-standing investment in social robotics for eldercare and companion applications (Aronsson, 2024) reflects a cultural receptivity to intelligent machines as caregiving allies. South Korea's high youth digital literacy and government-led AI ethics policies contribute to its relational readiness.

*Scandinavian Symmetry: Sweden*

Sweden (CS 8.6) exemplifies cultural alignment and trust architecture in AI integration. Its frameworks for transparency, ethical AI use, and emotional trust-building (Berman et al., 2024) enable consistent performance across all AFII pillars. With strong scores in symbolic receptivity and family-sensitive legal frameworks, Sweden provides a model for democratic and emotionally safe AI governance.

*United Kingdom and Germany: Ethical Infrastructure in Action*

The United Kingdom (CS 8.4) and Germany (CS 8.3) rank among the top six, thanks to their robust regulatory ecosystems, AI ethics institutions, and social policies that embed human values into technological systems. Their emphasis on informed consent, educational readiness, and caregiver inclusion supports sustainable AI integration in domestic contexts.

### 5.1.2 Diverging Middle Performers: Bridging Technological Strength and Human Readiness

Despite their AI dominance, China (CS 7.6) and the United States (CS 7.4) reveal gaps between innovation and emotional integration. While China scores high on infrastructure and deployment, challenges remain around symbolic use of AI in grief tech and afterlife rituals, raising ethical-cultural tensions (Cheng, 2025). Similarly, the U.S.—a leader in generative AI (Adolph et al., 2024)—underperforms in emotional caregiving integration, lacking comprehensive policies for AI in child-rearing or eldercare (Black, 2023).

The UAE (CS 6.6) and India (CS 6.0) show rising digital ambition but require focused interventions in emotional literacy, equity in caregiving access, and gender-sensitive AI design. India, in particular, is held back by regulatory fragmentation (Chatterjee, 2020) and persistent care-labor inequities (Bhallamudi, 2024), while the UAE is poised for progress through inclusive AI policies for disability and gender equity (Almufareh et al., 2024).

### 5.1.3 Lower Scorers and the Equity Imperative

Brazil (CS 5.2) and South Africa (CS 4.8) represent countries with immense potential but limited readiness. Their lower scores are shaped by structural constraints, including uneven access to AI infrastructure, lack of cultural trust frameworks, and limited ethical regulation. However, research (Aderibigbe et al., 2023) underscores the importance of building locally grounded AI narratives that resonate with indigenous values and community care practices. South Africa, for instance, has actively debated AI's role in child-rearing from a mental health standpoint (Andries et al., 2018), suggesting pathways for culturally adaptive caregiving AI.

## 5.2 Policy-Practice Gaps on AFI

As AI technologies increasingly permeate caregiving, emotional interaction, and familial domains, a critical question emerges: Do national policies genuinely support this symbolic and relational integration? This section examines the alignment between countries' AI Policy Penetration Scores—measured across the 10 aspects of the AI–Family Integration (AFI) framework—and their actual performance on the AI–Family Integration Index (AFII), which captures real-time implementation across AFI dimensions. By comparing these two indicators across 13 countries (refer to Table 27 and Figure 5), the analysis uncovers gaps between ethical intent and on-ground execution, highlighting where policy ambition may not yet translate into relationally responsible AI integration..

**Table 27: Policy-Practice Gaps: National AI Policy Penetration vs Real-time AFI**

| Country | National AI Policy Penetration based on AI aspects | | | Real-time AFI based on AI dimensions | | |
|---|---|---|---|---|---|---|
| | Composite Score | Rank | Penetration Power | Composite Score | Rank | Position |
| Brazil | 5.1 | 11 | Low | 5.2 | 12 | Low |
| China | 4.1 | 12 | Low | 7.6 | 8 | Moderate |
| France (EU) | 9.5 | 3 | High | 7.7 | 7 | Moderate |
| Germany (EU) | 9.5 | 4 | High | 8.3 | 6 | High |
| India | 6 | 10 | Moderate | 6 | 11 | Moderate |
| Japan | 9.6 | 2 | High | 8.7 | 3 | High |
| Singapore | 9.7 | 1 | High | 9.6 | 1 | High |
| South Africa | 3.6 | 13 | Low | 4.8 | 13 | Low |
| South Korea | 8.5 | 7 | High | 8.8 | 2 | High |
| Sweden (EU) | 8.6 | 6 | High | 8.6 | 4 | High |
| UAE | 6.8 | 9 | Moderate | 6.6 | 10 | Moderate |
| United Kingdom | 8.7 | 5 | High | 8.4 | 5 | High |
| United States | 7.9 | 8 | Moderate | 7.4 | 9 | Moderate |

Note: CS: Composite Score, High – CS (>8), Moderate – CS (8-6), Low – (CS <6)

Figure 5 visually contrasts the gap between national AI policy frameworks (marked by blue stars) and their real-time implementation or practice (marked by red circles) across countries. It highlights whether nations are effectively translating their AI policy ambitions into practice, as measured through AFI dimensions.

Table 27 complements this by presenting comparative data on composite scores for both AI policy penetration and real-time AFI performance, categorizing countries into high, moderate, or low penetration and practice levels. It enables a deeper understanding of alignment—or disconnect—between strategic intent and practical execution.

To create a Policy–Practice Matrix, classification criteria are applied to both policy and practice. In the assessment of both AI policy penetration and real-time practice on AI–Family Integration (AFI), countries are categorized as High, Moderate, or Low based on their composite scores. A country is classified as High when it attains a composite score greater than 8.0, indicating the presence of strong ethical policy frameworks or highly effective implementation of AFI components such as emotional, relational, and caregiving AI. Countries that score between 6.0 and 8.0 are placed in the Moderate category, reflecting efforts that are still in development. These countries typically demonstrate some integration of AFI dimensions, but not yet to a comprehensive or fully aligned extent. A Low classification is assigned to countries with a composite score below 6.0, which signifies limited policy readiness or real-time practice, particularly in areas like symbolic trust, digital personhood, and AI intimacy. The resultant matrix is shown in Figure 6, which maps countries across five quadrants based on the alignment between their AI policy intentions and real-time ethical implementation in AFI contexts.

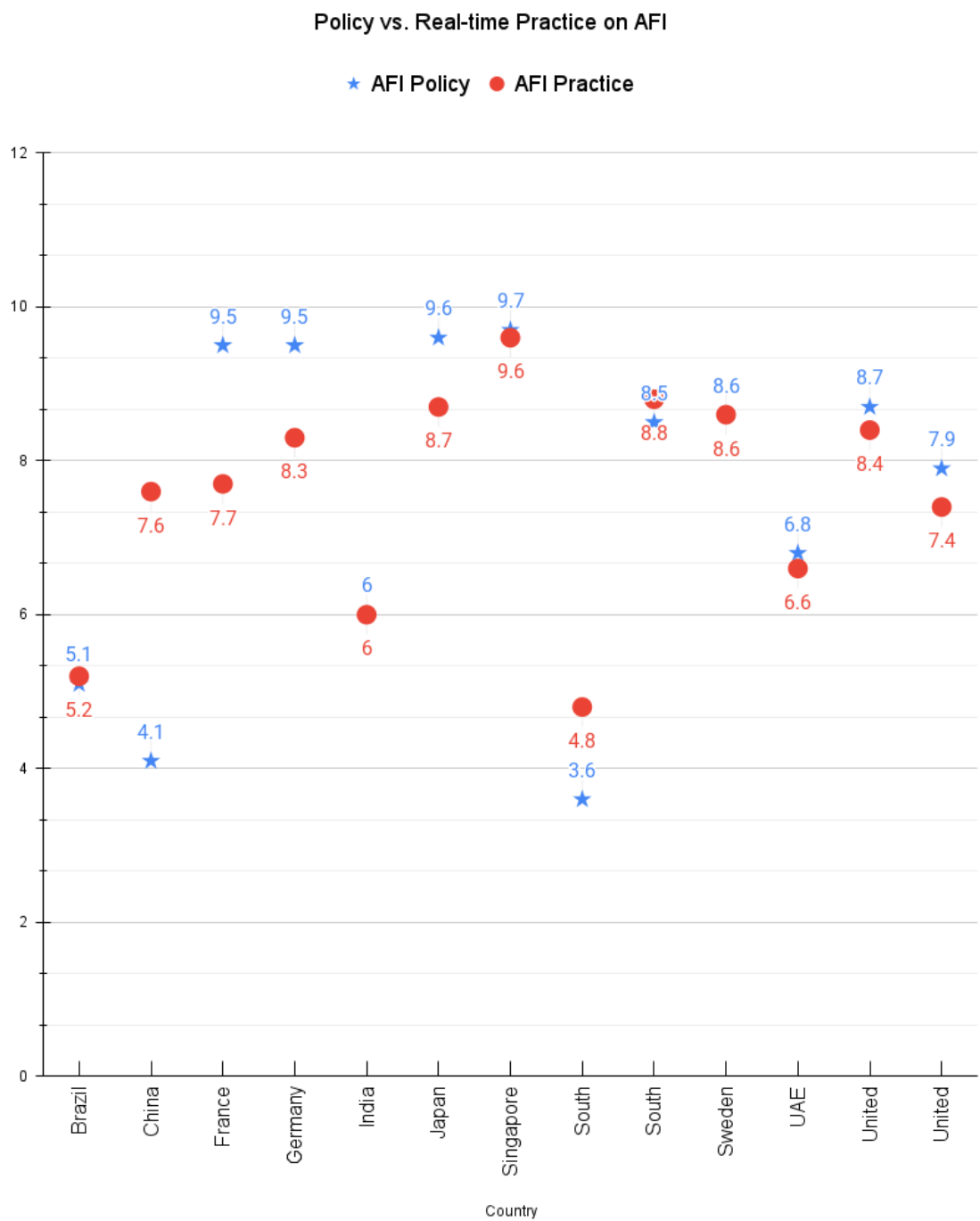

**Figure 5: Policy-Real-time Practice Gaps on AFI**

### 5.2.1   High Policy Penetration – High Practice on AFI (Integrated Ethical Leaders)

Singapore, Japan, United Kingdom, South Korea, and Sweden stand out as countries where national AI policies are not only comprehensive but also effectively implemented in real-world AI–Family Integration (AFI) contexts. These nations embed relational, emotional, and caregiving dimensions into their digital strategies and ethical frameworks. As a result, symbolic trust, AI caregiving, and digital personhood are not only regulated but reflected in practice through eldercare robotics, digital companions, and emotionally attuned AI tools. Their strong cultural alignment and policy clarity enable them to lead as true Integrated Ethical Leaders in AFI.

### 5.2.2 High Policy Penetration – Moderate Practice on AFI (Ethical Visionaries, Delayed Implementers)

France and Germany represent countries with strong policy intentions and ethical foundations—particularly guided by the EU's Trustworthy AI Guidelines and human-centered governance. However, their real-time AFII performance lags slightly behind their policy ranks. This suggests a more cautious rollout or regulatory bottlenecks in translating symbolic and emotional AI values into practice. These countries fall into the category of Ethical Visionaries, Delayed Implementers, with solid policy blueprints but slower implementation in domains like AI caregiving, AI intimacy, or kinship technologies.

### 5.2.3 Low Policy Penetration – High Practice on AFI (Unregulated Deployers)

China demonstrates a high real-time AFII score, driven by large-scale deployment of emotional robotics, AI assistants, and relational AI interfaces. However, these advancements have occurred with relatively minimal ethical governance regarding symbolic trust, digital personhood, or AI kinship. As such, China embodies the Unregulated Deployers quadrant—where practice significantly outpaces policy. While this approach accelerates innovation, it raises long-term concerns around accountability, equity, and cultural symbolism in AI-human relationships.

### 5.2.4 Moderate Policy Penetration – Moderate Practice on AFI (Ethical Middle Ground)

India, UAE, and the United States occupy a middle position in both policy penetration and real-time AFII readiness. These countries have launched national strategies and regulatory principles focused on responsible AI and inclusion. However, they often do not explicitly address symbolic or caregiving AI concerns, resulting in modest performance on AFI integration. As Ethical Middle Ground countries, they show potential for relational AI development but require clearer policy articulation and cross-sector implementation for meaningful progress.

### 5.2.5 Low Policy Penetration – Low Practice on AFI (Low Alignment, Low Readiness)

Brazil and South Africa currently score lowest in both AI policy penetration and AFII metrics. Their national AI strategies, while emerging, do not yet reflect sustained engagement with AI's symbolic, familial, or emotional dimensions. Without concrete policies or cultural emphasis on relational AI, these countries remain in the Low Alignment, Low Readiness quadrant—highlighting the need for foundational investment in both ethical framing and practical AFI deployment.

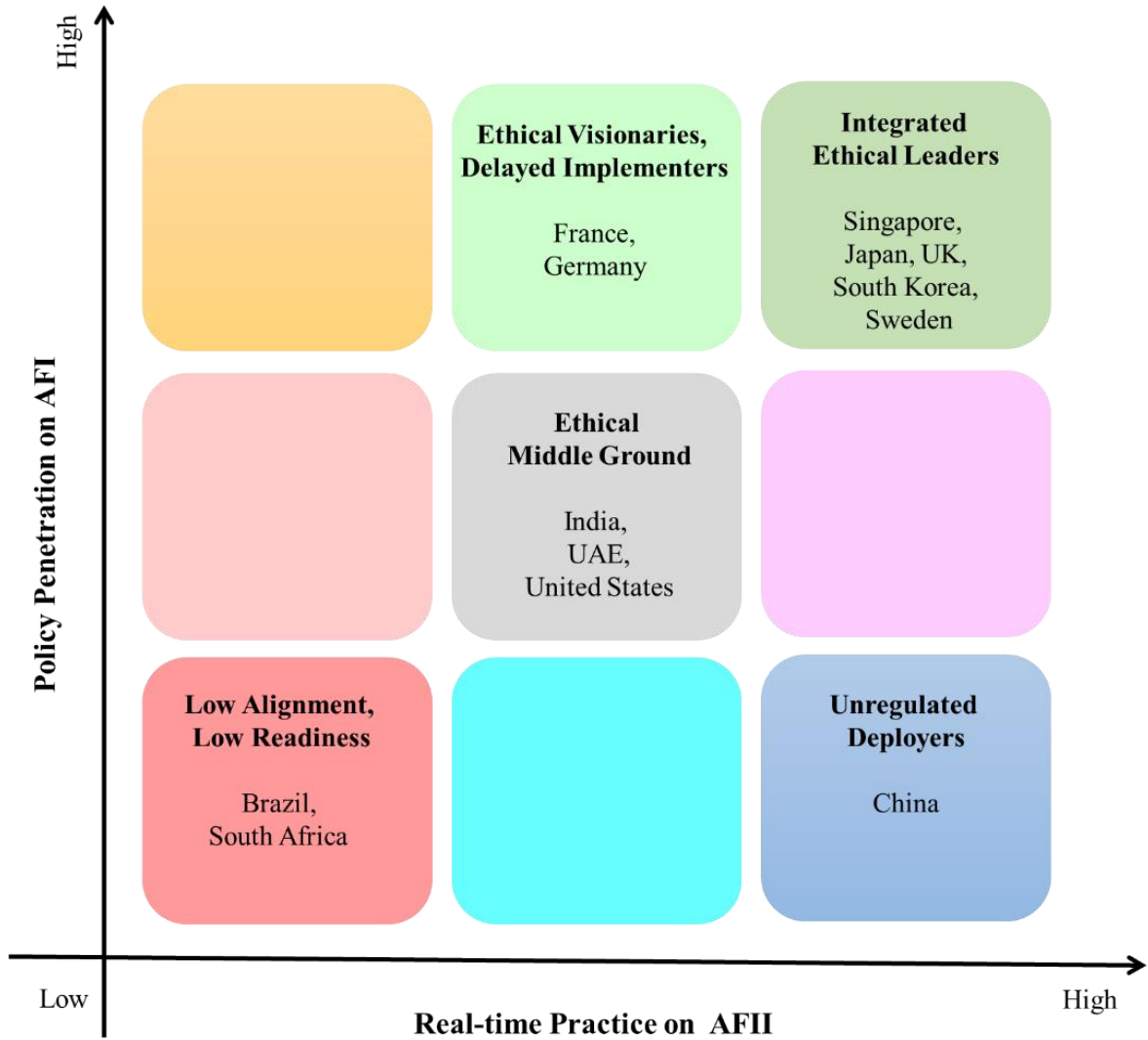

Figure 6: Policy-Practice Matrix

## 5.3    Contrasting Rankings: AI-Family Readiness vs AI Power

As artificial intelligence (AI) continues its global proliferation, two distinct yet complementary perspectives are emerging in the evaluation of national readiness. On one hand, AI Power—as measured by Stanford University's 2024 AI Index [77]—captures a country's influence in AI development, infrastructure, and economic contribution. On the other hand, the AI–Family Integration Index (AFII) introduced in this study highlights a nation's ethical, emotional, and cultural preparedness to integrate AI into caregiving, homes, and intimate social environments.

Table 28 and Figure 7 juxtapose a nation's rank on the AI–Family Integration Index (AFII)—which reflects emotional safety, symbolic trust, and inclusivity—with its position in the 2024 Stanford AI Index, which emphasizes AI infrastructure, research output, patents, governance, and economic indicators. The visual in Figure 7 clearly maps these differences, offering insight into which countries balance power with purpose.

| Country | AI-Family Integration Index | | 2024 AI Index (Power Rankings)* | |
|---------|----------------------------|---|--------------------------------|---|
| | Composite Score (CS) | AFII Rank | AI Rank (2024) | Strength for AI Rank |
| Singapore | 9.6 | 1 | 10 | Economy, diversity, responsible AI |
| South Korea | 8.8 | 2 | 7 | Governance, infrastructure, public opinion |
| Japan | 8.7 | 3 | 9 | Infrastructure, research, economy |
| Sweden | 8.6 | 4 | Not listed in Top 10 | N.A. |
| United Kingdom | 8.4 | 5 | 3 | R&D, education, responsible AI |
| Germany | 8.3 | 6 | 8 | Education, research, governance |
| France | 7.7 | 7 | 6 | Policy, education, infrastructure |
| China | 7.6 | 8 | 2 | R&D, economy, infrastructure |
| United States | 7.4 | 9 | 1 | Private investment, model development |
| UAE | 6.6 | 10 | 5 | Economic investment |
| India | 6.0 | 11 | 4 | R&D growth, tech economy |
| Brazil | 5.2 | 12 | Not listed in Top 10 | N.A. |
| South Africa | 4.8 | 13 | Not listed in Top 10 | N.A. |

**Table 28: Comparison between Country's AFII Rank Vs AI Index (2024)**

Note: Note: Countries not listed in the Stanford AI Index Top 10 include Sweden, Brazil, and South Africa.
*Source [77]

### 5.3.1 Key Highlights from Contrasting Rankings

- *Technological Dominance vs Human-Centered Readiness:*
- Despite ranking 1st and 2nd in global AI power, the United States (AFII Rank 9) and China (AFII Rank 8) demonstrate significant shortfalls in integrating AI within emotional, ethical, and caregiving frameworks. Their strengths in research, investment, and AI modeling are not matched by readiness in emotional safety, cultural receptivity, or symbolic trust. This gap illustrates a growing tension: AI expansion is outpacing ethical and relational integration.
- *Human-Centered Leaders Outperform in AFII:*

- Conversely, Singapore (AFII Rank 1, AI Rank 10) and Sweden (AFII Rank 4, Not in AI Top 10) are global exemplars in emotional intelligence, ethical governance, and inclusive design. Their high AFII scores show that AI leadership is not solely dependent on economic or computational strength. Instead, these countries offer blueprints for AI that prioritizes care, trust, and social well-being over sheer dominance.

- *Convergent Strengths in Policy and Practice:*
  Countries such as the United Kingdom (AFII Rank 5, AI Rank 3) and Germany (AFII Rank 6, AI Rank 8) display consistent excellence across both indexes. Their integration of research and development with ethical policy and emotional literacy positions them as balanced leaders in the future of AI. These nations are poised to shape international standards for responsible, emotionally aware AI ecosystems.

  *Dual Gaps in Emerging Economies:*
  Nations like India (AFII Rank 11, AI Rank 4), UAE (AFII Rank 10, AI Rank 5), Brazil (AFII Rank 12, Not in AI Top 10), and South Africa (AFII Rank 13, Not in AI Top 10) face twin challenges: building AI capability while also embedding emotional, ethical, and caregiving frameworks. While India excels in AI R&D, its care-labor imbalances and regulatory underdevelopment limit its relational readiness. These countries require early-stage investments in emotional-AI literacy, inclusivity, and symbolic trust-building

### 5.3.2 *Analytical Themes Emerging from the Contrast*

- *High-Tech Alone is Not Enough:*
  The data reveals a critical disconnect between AI power and family-level readiness. Nations like the U.S. and China dominate in AI research and commercial deployment, yet trail in symbolic trust, emotional design, and ethical care. As literature suggests (Etzioni & Etzioni, 2017; Abdollahi, 2023), technical leadership without social anchoring leads to brittle AI ecosystems vulnerable to public distrust and ethical misalignment.

- *Relational and Ethical Design as Strategic Assets:*
  Singapore and Sweden prove that emotional design, inclusive governance, and caregiving-centered frameworks can enhance global leadership even in the absence of overwhelming computational capacity. Their strength lies in recognizing that human trust, symbolic acceptance, and emotional safety are not optional extras—but core conditions for AI to thrive in human settings.

- *Convergent Leadership as the Future of Inclusive AI:*
  Germany and the UK exemplify convergent leadership, where technological innovation is underpinned by robust ethics, cultural trust, and relational frameworks. Their balanced standing across both indexes positions them to guide the global AI transition toward emotionally intelligent and ethically embedded applications.

- *A Call for South-South Collaboration:*
  For emerging nations, especially India, Brazil, and South Africa, the pathway forward lies in collaborative innovation. Their unique cultural diversity and large caregiving populations position them as laboratories for locally adaptive AI. Through partnerships with ethical AI leaders like Singapore and Sweden, they can accelerate the development of inclusive, trust-based AI systems that reflect their distinct societal needs.

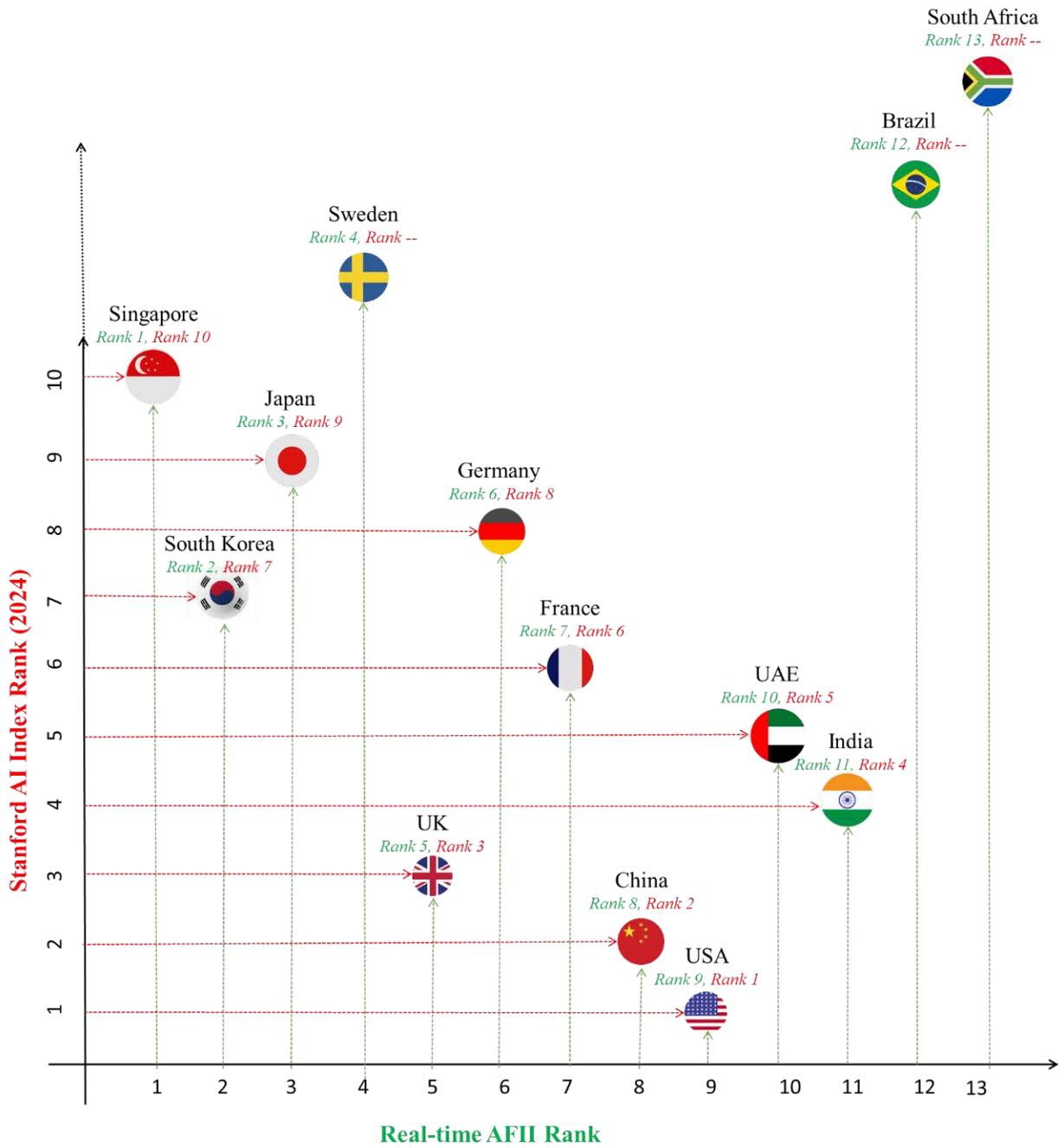

**Figure 7: Real-time AFII Rank vs Stanford AI Index Rank (2024)**

# 6. Discussions

As artificial intelligence (AI) increasingly enters caregiving, familial, and emotionally intimate domains, the central challenge for nations has evolved from innovation to integration with care. The AI–Family Integration Index (AFII) reveals that genuine readiness is not solely a matter of technological sophistication. Rather, it depends on a country's ability to align AI systems with emotional intelligence, symbolic trust, and cultural resonance.

This section synthesizes findings from the AFII to offer cross-cutting insights, strategic pathways, and visionary directions for building emotionally intelligent AI futures that are ethical, inclusive, and context-sensitive.

This discussion affirms the successful fulfilment of the study's four research objectives and corresponding four research questions, as outlined in the updated framework. The preceding sections established the conceptual foundation, constructed the AFII, applied it across 13 countries, and extracted comparative insights. What follows is a thematic, visionary synthesis structured to shape policy, innovation, and global cooperation toward emotionally aware and relational AI futures.

## 6.1 Cross-Cutting Themes and Strategic Calls to Action

- ***Emotional Literacy and Trust Design: Core Pillars of Integration***

  Once viewed as soft or secondary, emotional intelligence is now a foundational requirement for AI's safe and meaningful integration into human spaces. Countries that score high in AFII dimensions such as emotional literacy (ATYEEL) and trust design (SNST, EASD) show a deep understanding of the need for emotionally congruent, context-aware AI interactions—especially in environments involving children, elders, and caregivers. This insight demands strategic investment in AI ethics co-developed with affective computing experts, emotional-AI literacy embedded in school curricula, and emotional trust audits integrated into national AI evaluations. In caregiving settings, emotional intelligence is no longer optional—it is essential for AI to be accepted, trusted, and safe.

  "The integration of artificial intelligence into caregiving, emotional support, and family life signals a pivotal shift in both technological development and societal structure. AI must be measured not just by what it can do, but by how it is trusted to belong in our most intimate spaces."

  This statement underscores a central insight of the study: that readiness for AI is no longer about technical capacity alone—it is about trust, belonging, and emotional legitimacy. Framing this as a benchmark for "Relational AI Maturity" may help policymakers adopt a clearer lexicon for non-technical AI readiness.

- ***Cultural Adaptability, Symbolic Receptivity & Local Inclusivity***

Top-scoring nations demonstrate not only strong policy frameworks, but alignment with symbolic values, cultural norms, and family philosophies. This alignment enhances public trust and accelerates adoption.

Strategic imperatives include:

o Culturally co-designed AI systems reflecting linguistic and familial diversity;
o Participation of caregivers, elders, and underrepresented communities in design and governance;
o Policies that bridge the urban-rural AI divide and promote symbolic resonance.

- ***Ethical Governance and Policy Clarity: The Missing Infrastructure***

  The AFII reveals a lag between ethical intent and relational implementation. While many countries articulate AI ethics principles, few possess enforceable frameworks for emotional safety, dynamic consent, or caregiving-specific regulation.

  Policy needs include:
  o Cross-sector AI commissions (health, education, family welfare);
  o Emotional safety certification standards for domestic AI;
  o Global ethical standards for AI in caregiving and companionship roles.

These themes collectively lay the groundwork for a future in which AI integration is assessed through emotional legitimacy, symbolic trust, and cultural fit—not merely by computational power, fulfilling of Objective 1 and respond to Research Question 1.

## 6.2 Toward an Inclusive, Emotionally Intelligent AI Future

The AFII is not merely a benchmarking tool—it is a strategic compass. High-ranking nations (e.g., Singapore, South Korea, Japan) must export care-centric AI protocols, lead global ethics dialogues, and support South–South collaboration. Meanwhile, middle- and lower-ranking nations can harness hybrid innovation, merging global best practices with indigenous values to design culturally resonant, trust-based AI systems.

This section reframes emotional innovation not as a lag but as a competitive advantage. Emotional richness—particularly in caregiving traditions—can offer future leadership potential.

The true metric of AI success in family life will not be devices or data—but the depth of empathy, trust, and inclusion it fosters.

These insights confirm the fulfilment of Objective 2 and address Research Question 2

## 6.3 Visionary Suggestions: From Emotional Literacy to Symbolic AI Integration

The AFII findings reflect a transformative shift in how AI readiness should be understood—not merely as a function of computational capability or innovation output, but as a nation's ability to emotionally, ethically, and culturally embed AI within caregiving and familial domains. High-scoring countries like Singapore, South Korea, and Japan exemplify this readiness. Singapore's top composite score of 9.6 results from a seamless alignment between emotionally intelligent AI systems and caregiving frameworks, including eldercare and early education.

What makes the top performers visionary is not just their policy clarity or infrastructure, but their recognition of symbolic trust and emotional authority as critical dimensions of readiness. Sweden and the United Kingdom further validate this premise by representing democratic and transparent governance models that elevate emotional literacy and inclusiveness as strategic national assets.

Conversely, middle- and lower-scoring countries like India, Brazil, and South Africa possess a distinct advantage: cultural and emotional depth. The richness of their caregiving traditions, spiritual philosophies, and community narratives makes them fertile grounds for localized, trust-based AI development.

These nations are not merely late adopters—they are potentially emotional innovators capable of setting new paradigms for AI grounded in social empathy and symbolic resonance.

These observations support the fulfilment of Objective 3 and address Research Question 3.

## 6.4    Strategic Recommendations: Redesigning AI Maturity through Relational Integration

While China and the United States lead in AI power rankings, their AFII scores highlight significant gaps in symbolic trust, emotional integration, and ethical grounding. Their models, often progress-driven, have yet to invest deeply in the relational foundations necessary for AI to be fully trusted in caregiving and domestic roles.

To address this, emotional literacy, symbolic resonance, and caregiving ethics must become central to AI maturity. Countries should prioritize decentralizing policy to empower regional frameworks, like India's "Bharat AI-FAMILY Mission" or Brazil's potential Afro-Brazilian-led caregiving labs. Equally important is the extension of AI development beyond urban areas, with trials in semi-urban regions like Xi'an or Raipur to reflect everyday relational realities.

Education must include emotional-AI literacy, combining empathy training and ethical reasoning with technical fluency. National AI strategies should integrate emotional trust tools, such as safety dashboards and relational behavior registries, to monitor and validate emotionally intelligent performance in sensitive use cases.

These recommendations are shaped by the comparative findings and reflect the fulfilment of Objective 4 while responding directly to Research Question 4.

## 6.5    Cultural and Religious Diversity: A Catalyst for Global AI Leadership

The AFII underscores that cultural and religious diversity—especially when paired with high population—offers a competitive edge in the global AI economy. Countries like India, Brazil, and South Africa hold vast emotional vocabularies that make relational AI more adaptable, trustworthy, and culturally embedded.

Rather than hinder innovation, diversity fosters emotional realism, enhances symbolic trust, and enables scalable inclusion. AI systems designed with ethical pluralism, multilingual sensitivity, and region-specific emotional grounding are more likely to resonate with real users.

This reframing encourages policymakers to view diversity not as a limitation but as emotional capital. It enables relational AI to scale in education, health, parenting, and eldercare—transforming social value into economic growth.

To strengthen this section, examples could be included on how religious or spiritual traditions (e.g., animism in Japanese robotics, Ubuntu in African technology) have shaped technological trust.

This exploration further supports Objective 4 and expands on the implications identified in response to Research Question 4.

### 6.6 Actionable Plans: Building Emotionally Intelligent Ecosystems through Infrastructure and Participation

- *From Vision to Implementation: Grounding Emotional AI in Public Infrastructure:*
  To translate emotional readiness into development outcomes, countries must invest in emotionally intelligent ecosystems supported by inclusive infrastructure, participatory governance, and trust-building mechanisms. High-population, culturally plural societies are ideal launchpads for relational AI.

- *Institutional Trust: Embedding Emotional Safety in Governance Frameworks:*
  Embedding emotional trust into national AI evaluations through dashboards, registries, and community feedback ensures accountability and alignment with cultural expectations. Nations must prioritize emotional safety in caregiving, grief support, and intergenerational AI interactions.

- *Innovation from the Margins: Scaling Grassroots Emotional-AI Labs:*
  Decentralized co-design labs in tribal, rural, and underserved regions should be treated as innovation epicenters. Governments must fund and scale emotionally diverse AI systems through local partnerships that reflect the real-life relational and ethical worldviews of the population.

- *National Platforms for All: Delivering Relational AI as a Public Good:*
  Publicly funded, multilingual AI-family platforms can provide grief bots, parenting assistants, and eldercare services as essential emotional infrastructure. Examples like India's "Bharat AI-FAMILY Channel" or Brazil's "Família+AI" can democratize access to emotional intelligence technologies.

- *Symbolic Legitimacy through Culture: Building Trust via Public Imagination:*
  Celebrity-led campaigns rooted in cultural storytelling can enhance emotional acceptance. Initiatives like "AI as Family: Together for Tomorrow" in India or "Afeto Digital" in Brazil allow communities to emotionally relate to AI as a caregiving ally.

- *Emotion as Economy: Turning Diversity into Scalable Growth:*
  High-population countries can lead the AI care economy by translating cultural richness into service innovation. Emotional AI is not only socially transformative—it is economically catalytic.

- ***The Triad of Transformation: Infrastructure, Participation, and Emotional Design:*** Nations that invest in inclusive infrastructure, participatory design, and emotional trust will not only improve AFII performance but also emerge as global leaders in relational AI.

These pathways extend the fulfilment of Objective 4 and deepen the response to Research Question 4.

## 6.7 Global Frameworks for AFI Futures

To ensure that AI systems integrated into family and caregiving environments are ethical, emotionally intelligent, and culturally aligned, a coordinated global approach is necessary. Nations must move beyond isolated AI strategies and work toward collective frameworks that support relational AI development at an international scale. This section proposes three key initiatives to shape the future of AI–Family Integration (AFI) through cross-national collaboration and shared governance.

- ***A Global AFI Charter:***
  A Global AI–Family Integration Charter, developed under international organizations such as UNESCO or OECD, would provide foundational guidelines for embedding AI into emotionally and socially sensitive domains. This charter would establish:

  o Cultural Caregiving Norms: AI should be designed with respect for diverse caregiving traditions, from multigenerational family structures to community-driven child-rearing models.

  o Affective Consent Standards: AI systems interacting with children, elders, and vulnerable populations must follow stringent consent mechanisms that account for emotional safety and ethical relationality.

  o Symbolic Trust Metrics: AI tools used in family settings should meet predefined trustworthiness benchmarks, ensuring they enhance relational bonds rather than disrupt them.

- ***Open AFI Data Commons:***
  A global repository of AI–Family Integration data, built collaboratively by researchers, governments, and local communities, would enable the development of culturally adaptive relational AI models. This initiative would:

  o Collect multilingual, multi-generational, and multi-cultural caregiving interactions, ensuring relational AI systems reflect real-world emotional complexity.

  o Promote community-driven data annotation, preventing Western-biased interpretations of emotional intelligence and family dynamics.

  o Ensure public access for ethical AI development, particularly for startups, educators, and social impact organizations in the Global South.

- ***Global South AFI Leadership Coalition:***
  A strategic alliance of Global South nations—led by India, Brazil, China, South Africa, and Indonesia—would co-develop standards and lead pilot initiatives for AI integration into caregiving and family life. This coalition would:

  o Co-create culturally responsive AI guidelines, ensuring AI caregiving aligns with regional emotional norms and caregiving values.

  o Cross-test relational AI models in diverse family structures, from extended family households in India to community-care models in Africa.

  o Develop and export ethically aligned AI tools, allowing nations to take leadership in AI caregiving, elder support, and emotional AI governance on a global scale.

By implementing these global frameworks, AI–Family Integration (AFI) can evolve into a universally trusted, ethically responsible, and culturally intelligent paradigm—ensuring that AI strengthens, rather than replaces, the deep emotional ties that define human caregiving. These global frameworks extend the fulfilment of Objective 5 and strengthen the response to Research Question 5.

## 6.8    Limitations and Future Direction

The current scope of the AFII is limited to a relatively small group of countries, which restricts its global applicability; future research should therefore expand the index to include more diverse and underrepresented nations, especially from regions such as the Middle East, Central Asia, and Sub-Saharan Africa, to build a truly global picture of relational AI readiness. Several AFII dimensions—particularly symbolic trust and emotional authority—rely on interpretive assessment methods that are sensitive to cultural variation and subjectivity; future studies should develop standardized and culturally adaptable indicators to enable more consistent and valid cross-national comparisons.

The framework currently lacks direct input from AI users, such as caregivers, elders, and children, which limits its grounding in lived experience; future research should employ participatory approaches, including ethnographic case studies and user diaries, to integrate the emotional and relational realities of everyday AI interaction.

There is also an absence of longitudinal data on how relational AI systems affect psychological wellbeing, trust-building, and emotional resilience over time; to address this, future research should design long-term studies that track the emotional and symbolic impact of AI within households and caregiving environments.

The lack of global consensus on emotional safety protocols and affective consent makes benchmarking across countries difficult; future efforts should focus on co-developing international ethical standards and emotional safety frameworks, potentially through interdisciplinary collaborations involving ethicists, regulators, and cultural scholars.

While the AFII discusses cultural and philosophical receptivity qualitatively, these dimensions are not yet quantified; future studies should explore proxies such as the presence of AI in cultural narratives, spiritual discourse, or educational storytelling, to systematically measure symbolic and emotional integration.

In addition, the index currently assesses existing conditions but does not measure a society's adaptive capacity to relational AI over time; future iterations should incorporate indicators of policy flexibility, social learning, and resilience to reflect evolving AI-human dynamics. Methodological limitations are also acknowledged in the Research Methodology section.

To move from a prototype to a globally representative framework, the AFII must now enter a phase of iterative co-creation. This includes interdisciplinary engagement with cultural scholars, technologists, public health experts, and AI users themselves. As AI becomes increasingly embedded in emotional and caregiving domains, the need for a responsive, inclusive, and globally valid index is more urgent than ever. The future of relational AI will require not only better data—but deeper dialogue, cultural humility, and shared ethical vision. The AFII can serve as both a mirror and a map—reflecting present gaps while charting the path toward a more emotionally intelligent AI future.

## 7.    Conclusion

The integration of artificial intelligence into caregiving, emotional support, and family life marks a profound shift in both technological progress and societal expectations. As AI systems increasingly enter spaces of intimacy—helping people parent, grieve, educate, and care—the traditional understanding of "AI readiness" must evolve. Existing global benchmarks, such as the Stanford AI Index, provide valuable insights into infrastructure and innovation. However, they fall short in capturing the emotional, ethical, and cultural conditions that shape AI's acceptance in the relational fabric of everyday life.

To address this critical gap, the AI–Family Integration Index (AFII) was introduced as a multidimensional framework that evaluates not only what AI systems can do, but how they are emotionally trusted, symbolically embedded, and culturally resonant in caregiving and domestic environments. Through ten ethically and emotionally grounded dimensions—including emotional safety, symbolic trust, caregiving equity, and youth exposure—the AFII redefines AI maturity as a function of empathy, relational alignment, and inclusive design.

The application of the AFII reveals a changing global AI landscape. Nations such as Singapore, South Korea, Japan, and Sweden emerge as leaders not solely due to technological capacity, but because they integrate emotional literacy, anticipatory governance, and care-centered values into their national AI strategies. These countries exemplify models where AI is not only operationally effective, but emotionally legitimate and socially trusted. In contrast, countries like the United States and China—though dominant in infrastructure and research—exhibit a disconnect between innovation and relational maturity, often sidelining caregiving ethics, emotional design, and symbolic integration in their AI deployments.

A second critical insight is the persistence of a policy–practice gap. While many nations articulate ambitious ethical frameworks, few have translated those visions into emotionally safe and culturally adaptive AI systems within real caregiving settings. For instance, France and Germany emphasize human-centered AI in official discourse but lag in deploying relational tools at scale. Meanwhile, China and South Korea have advanced affective robotics but often lack robust mechanisms for emotional consent and symbolic legitimacy. These disjunctions highlight the need for emotional infrastructure, participatory governance, and culturally situated safeguards to ensure AI systems are not only functional, but trusted to cohabit our most intimate spaces.

The study also draws attention to the limitations of traditional AI benchmarking tools. While indices like the Stanford AI Index remain important for tracking technical growth, they fail to measure the emotional realism, caregiving adaptability, and symbolic resonance that define AI's role in daily life. The AFII offers a corrective and complementary lens, reframing readiness in terms of relational ethics and emotional integration, rather than pure computational performance.

At the same time, emerging economies—such as India, Brazil, and South Africa—despite scoring lower on technical infrastructure, possess rich caregiving traditions, spiritual philosophies, and communal family models. These emotional and cultural assets position them as potential emotional innovators. With targeted investment in emotional-AI education, localized interface design, and grassroots co-design labs, they could set new global precedents for culturally embedded, trust-based AI.

For policymakers, the findings carry important implications. National AI strategies must embed emotional literacy and caregiving ethics as foundational components of AI education and development. AI systems should reflect the diverse emotional vocabularies and caregiving norms of pluralistic societies. Community-based innovation—particularly in rural, tribal, and underserved regions—must be scaled and funded. Most importantly, AI governance must become interdisciplinary and culturally grounded, drawing on the perspectives of caregivers, technologists, ethicists, educators, and cultural scholars to ensure emotionally intelligent futures.

In addition to national strategies, global frameworks will play a pivotal role in guiding equitable and emotionally safe AI integration. Initiatives such as a Global AI–Family Integration Charter, an Open AFI Data Commons, and a Global South AFI Coalition can foster international collaboration, harmonize emotional safety standards, and support context-sensitive innovation. These frameworks ensure that relational AI development is not siloed by region or ideology, but anchored in shared ethical principles, cross-cultural dialogue, and globally accessible emotional infrastructure.

AFII will evolve in future iterations to include a broader set of nations, participatory user data, and longitudinal indicators that track AI's emotional and symbolic impacts over time—ensuring the framework remains globally relevant and deeply grounded in lived human experiences.

Ultimately, the AFII redefines AI readiness as a question of care—not just capability. The true benchmark for AI is not whether it can automate tasks or simulate intelligence, but whether it can be trusted to share space in the emotional and symbolic dimensions of human life.

The time to integrate care into AI governance is now. Nations that lead with empathy, inclusion, and cultural imagination will not only shape technology—but chart the moral course of the digital age.